\documentclass[journal]{IEEEtran}

\usepackage[ruled,linesnumbered]{algorithm2e}
\usepackage{enumerate}
\usepackage{epsfig}
\usepackage{graphicx}
\usepackage{amssymb}
\usepackage{url}
\usepackage{subfigure}
\usepackage{wrapfig}
\usepackage{paralist}
\usepackage{xcolor}
\usepackage{booktabs} 
\usepackage{tikz}
\usepackage{pifont}

\newcommand{\cmark}{\ding{51}}%
\newcommand{\xmark}{\ding{55}}%

\newcommand{\clean}{\textit{clean}}
\newcommand{\dirty}{\textit{dirty}}
\newcommand{\start}{\textit{start}}

\begin{document}

\title{Secure Management of Low Power Fitness Trackers}

\author{
   \IEEEauthorblockN{
   Mahmudur Rahman\IEEEauthorrefmark{1},
   Bogdan Carbunar\IEEEauthorrefmark{1}, Member IEEE,
   Umut Topkara\IEEEauthorrefmark{2}}, Member IEEE,\\
   \IEEEauthorblockA{\IEEEauthorrefmark{1}Florida International University, Miami, FL}\\
   \IEEEauthorblockA{\IEEEauthorrefmark{2}JW Player, New York, NY}\\
   \thanks{A preliminary concise version of this paper appears in IEEE ICNP 2014.}
   \thanks{This research was supported in part by DoD grant W911NF-13-1-0142 and NSF grant EAGER-1450619.}
}

\maketitle

\begin{abstract}

The increasing popular interest in personal telemetry, also called the
Quantified Self or ``lifelogging'', has induced a popularity surge for wearable
personal fitness trackers. Fitness trackers automatically collect sensor data
about the user throughout the day, and integrate it into social network
accounts.  Solution providers have to strike a balance between many
constraints, leading to a design process that often puts security in the back
seat. Case in point, we reverse engineered and identified security
vulnerabilities in Fitbit Ultra and Gammon Forerunner 610, two popular and
representative fitness tracker products. We introduce FitBite and GarMax, tools
to launch efficient attacks against Fitbit and Garmin.

We devise SensCrypt, a protocol for secure data storage and communication, for
use by makers of affordable and lightweight personal trackers. SensCrypt
thwarts not only the attacks we introduced, but also defends against powerful
JTAG Read attacks. We have built Sens.io, an Arduino Uno based tracker
platform, of similar capabilities but at a fraction of the cost of current
solutions. On Sens.io, SensCrypt imposes a negligible write overhead and
significantly reduces the end-to-end sync overhead of Fitbit and Garmin.

\end{abstract}


\section{Introduction}

\begin{figure}
\centering
\subfigure[]
{\label{fig:system:main}{\includegraphics[width=1.7in]{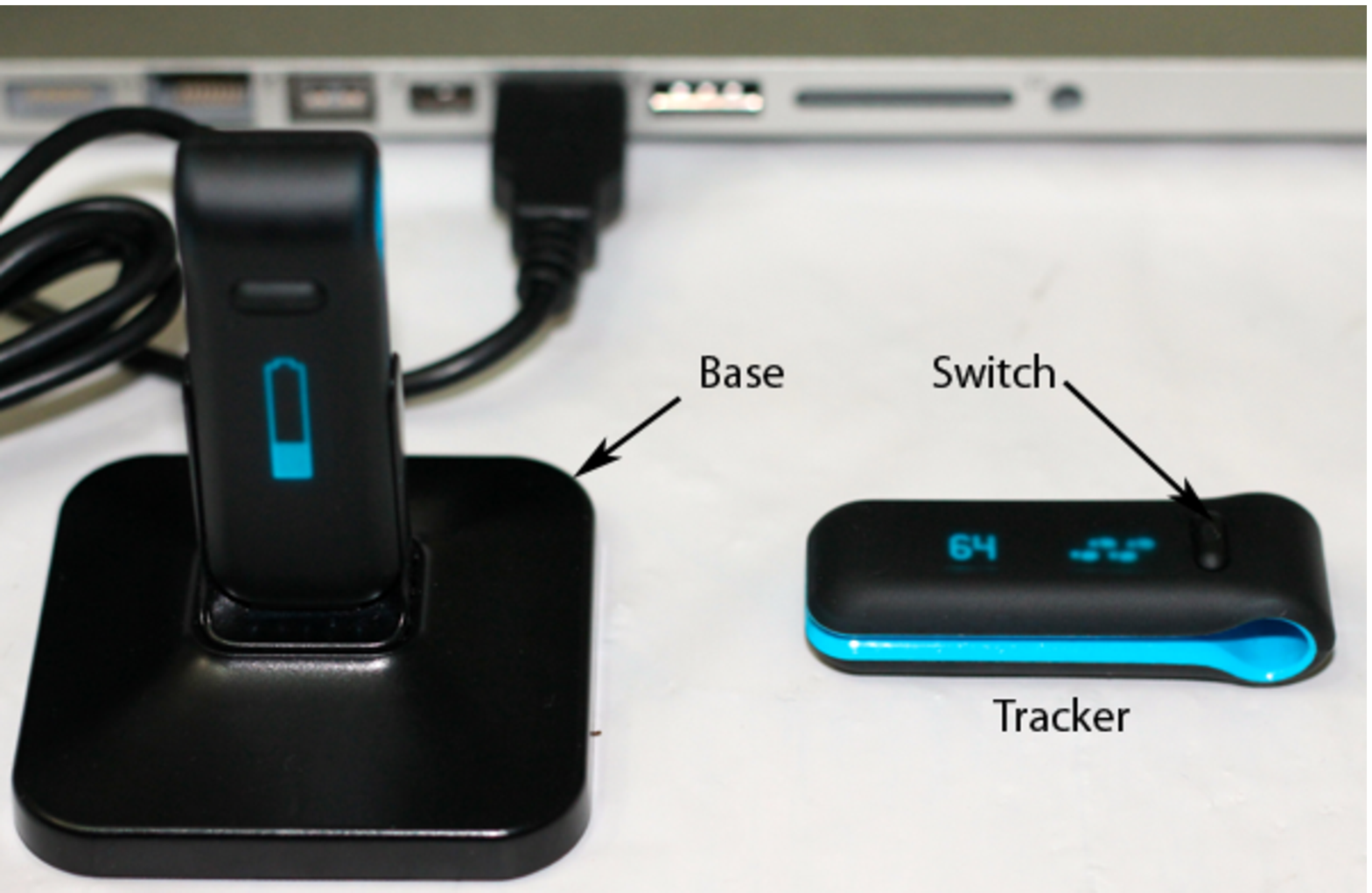}}}
\subfigure[]
{\label{fig:system:mainGarmin}{\includegraphics[width=1.7in]{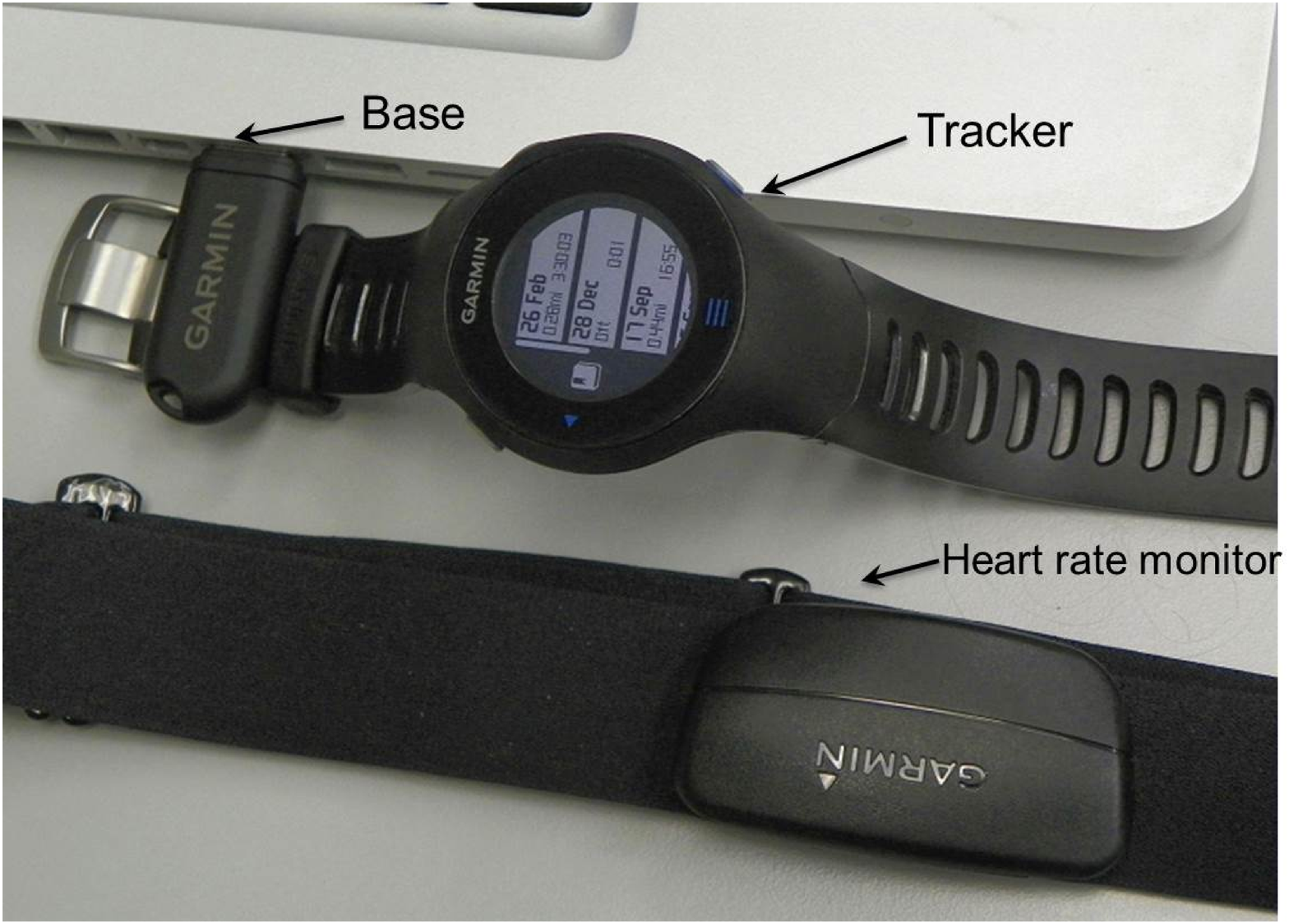}}}
\caption{System components:
(a) Fitbit: trackers (one cradled on the base), the base
(arrow indicated), and a user laptop. The arrow pointing to the tracker shows
the switch button, allowing the user to display various fitness data.
(b) Garmin: trackers (the watch), the base(arrow indicated), and a user laptop.}
\end{figure}

Wearable personal trackers that collect sensor data about the wearer, have long
been used for patient monitoring in health care.  Holter
Monitors~\cite{HolterM}, with large and heavy enclosures, that use tapes for
recording, have recently evolved into affordable personal fitness trackers
(e.g.,~\cite{Nike+}). Recently, popular health centric \textit{social sensor
networks} have emerged. Products like Fitbit~\cite{Fitbit}, Garmin
Forerunner\cite{Forerunner} and Jawbone Up~\cite{Jawbone} require users to
carry wireless trackers that continuously record a wide range of fitness and
health parameters (e.g., steps count, heart rate, sleep conditions), tagged
with temporal and spatial coordinates. Trackers report recorded data to a
providing server, through a specialized wireless base, that connects to the
user's personal computer (see Figures~\ref{fig:system:main}
and~\ref{fig:system:mainGarmin}). The services that support these trackers
enable users to analyze their fitness trends with maps and charts, and share
them with friends in their social networks.

All happening too quickly both for vendors and users alike, this data-centric
lifestyle, popularly referred to as the Quantified Self or ``lifelogging'' is
now producing massive amounts of intimate personal data. For instance,
BodyMedia~\cite{BodyMedia} has created one of the world's largest libraries of
raw and real-world human sensor data, with 500 trillion data
points~\cite{BMData}. This data is becoming the source of privacy and security
concerns: information about locations and times of user fitness activities can
be used to infer surprising information, including the times when the user is
not at home~\cite{PleaseRobMe}, and company organizational
profiles~\cite{TKS13}.

We demonstrate vulnerabilities in the storage and transmission of personal
fitness data in popular trackers from Fitbit~\cite{Fitbit} and
Garmin~\cite{Forerunner}. Vulnerabilities have been identified for similar
systems, including pacemakers (e.g., Halperin et al.~\cite{HHBRCDMFKM08}) and
glucose monitoring and insulin delivery systems (e.g., Li et.
al.~\cite{Insulin}). The differences in the system architecture and
communication model of social sensor networks enable us to identify and exploit
different vulnerabilities.

We have built two attack tools, FitBite and GarMax, and show how they 
inspect and inject data into nearby Fitbit Ultra and Garmin
Forerunner trackers. The attacks are fast, thus practical even during brief
encounters.  We believe that, the vulnerabilities that we identified in the
security of Fitbit and Garmin are due to the many constraints faced by solution
providers, including time to release, cost of hardware, battery life, features,
mobility, usability, and utility to end user.  Unfortunately, such a
constrained design process often puts security in the back seat.

To help address these constraints, in this paper we introduce SensCrypt, a
protocol for secure fitness data storage and transmission on lightweight
personal trackers. We leverage the unique system model of social sensor
networks to encode data stored on trackers using two pseudo-random values, one
generated on the tracker and one on the providing server. This enables
SensCrypt, unlike previous work~\cite{HHBRCDMFKM08,RCHBC09}, to protect not
only against inspect and inject attacks, but also against attackers that
physically capture and read the memory of trackers.  SensCrypt's hardware and
computation requirements are minimal, just enough to perform low-cost symmetric
key encryption and cryptographic hashes.  SensCrypt does not impose storage
overhead on trackers and ensures an even wear of the tracker storage, extending
the life of flash memories with limited program/erase cycles.

SensCrypt is related to Dabinder (Naveed et al.~\cite{NZDWG14}), an Android
level defense. Dabinder generates and enforces secure bonding policies between
a device and its official app, to prevent external device mis-bonding attacks
for Bluetooth enabled Android health/medical devices. SensCrypt is built for a
different platform and also, unlike Dabinder, minimizes the role played by
the base.


SensCrypt is applicable to a range of sensor based platforms, that includes a
large number of popular fitness~\cite{Fitbit,Forerunner,Jawbone,MotoActv,Basis}
and home monitoring solutions~\cite{Nest,WeMo,Mother}, as well as scenarios
where the sensors need to be immobile and operable without network connectivity
(e.g., infrastructure, traffic, building and campus monitoring solutions).  In
the latter case, the bases through which the sensors sync with the webserver
are mobile, e.g., smartphones of workers, who may become proximal to the
sensors with the intention of data collection or as a byproduct of routine
operations.

We have developed Sens.io, a \$52 tracker platform built on Arduino Uno, of
similar capabilities with current solutions. On Sens.io, SensCrypt (i) imposes
a 6ms overhead on tracker writes, (ii) reduces the end-to-end overhead of data
uploads to 50\% of that of Fitbit, and (iii) enables a server to support large
volumes of tracker communications. In conclusion, the contributions of this
paper are the following:

\begin{compactitem}

\item
Reverse engineer the semantics of the Fitbit Ultra and Garmin Forerunner
communication protocol.  [Section~\ref{sec:model:reverse}].

\item
Build FitBite and GarMax, tools that exploit vulnerabilities in the design of
Fitbit and Garmin to implement several attacks in a timely manner
[Section~\ref{sec:attacks}].

\item
Devise SensCrypt, a secure solution that imposes no storage overhead on
trackers and requires only computationally cheap operations.
[Section~\ref{sec:solution}] Show that SensCrypt protects even against invasive
attackers, capable of reading the memory of captured trackers
[Section~\ref{sec:analysis}].

\item
Implement Sens.io, a tracker platform, of similar capabilities with existing
popular solutions but at a fraction of the cost
[Section~\ref{sec:implementation}]. Show that SensCrypt running on Sens.io is
very efficient [Section~\ref{sec:evaluation}]

\end{compactitem}

While SensCrypt's defenses may not be immediately adopted by existing
products~\footnote{We have contacted Fitbit and Garmin with our results. While
interested in the security of their users, they have declined collaboration.},
this paper provides a foundation upon which to create, implement and test new
defensive mechanisms for future tracker designs.

\section{System Model, Attacker and Background}
\label{sec:model}

\subsection{System Model}
\label{sec:model:system}

We consider a general system consisting of tracker devices, base stations and
an online social network.  We exemplify the model components using Fitbit
Ultra~\cite{Fitbit} and Garmin Forerunner~\cite{Forerunner}, two popular health
centric social sensor networks (see Figures~\ref{fig:system:main}
and~\ref{fig:system:mainGarmin}).  For simplicity, we will use ``Fitbit'' to
refer the Fitbit Ultra and ``Garmin'' to denote the Garmin Forerunner 610
solution.

\noindent
{\bf The tracker.}
The \textit{tracker} is a wearable device that records, stores and reports a
variety of user fitness related metrics. We focus on the following
trackers:

\begin{compactitem}

\item
{\bf The Fitbit tracker} measures the daily steps taken, distance traveled,
floors climbed, calories burned, the duration and intensity of the user
exercise, and sleep patterns.  It consists of four IC chips, (i)
a MMA7341L 3-axis MEMS accelerometer, (ii) a MEMS altimeter to count the number
of floors climbed and (iii) a MSP 430F2618 low power TI MCU consisting of 92 KB
of flash and 96 KB of RAM. The user can switch between displaying different
real-time fitness information on the tracker, using a dedicated hardware
\textit{switch} button (see the arrow pointing to the switch in
Figure~\ref{fig:system:main}).

\item
{\bf The Garmin tracker} records data at user set periodic intervals (1-9
seconds). The data includes a timestamp, exercise type, average speed, distance
traveled, altitude, start and end position, heart rate and calories burned
during the past interval. The tracker has a heart rate monitor (optional) and a
12 channel GPS receiver with a built-in SiRFstarIII antenna.  that enables the
user to tag activities with spatial coordinates.

\end{compactitem}

\noindent
Both Fitbit and Garmin trackers have chips supporting the ANT protocol, with a
15ft transmission range for Fitbit and 33ft for Garmin.  Each tracker has a
unique id, called the \textit{tracker public id} (TPI). Trackers also store
profile information of their users, including age, gender and physiological
information such as height, weight and gait information.

\noindent
{\bf The base and agent module.}
The base connects with the user's main computing center (e.g., PC, laptop) and
with trackers within transmission range (15ft for Fitbit and 33ft for
Forerunner) over the ANT protocol. The user needs to install an ``agent
module'', a software provided by the service provider (Fitbit, Garmin) to run
on the base. The agent and base act as a bridge between the tracker and the online
social network. They upload information stored on the tracker to its user
account on the webserver, see Figures~\ref{fig:system:main}
and~\ref{fig:system:mainGarmin} for system snapshots.

\noindent
{\bf Tracker to base pairing}.
Fitbit trackers communicate to any base in their vicinity. However, tracker
solutions like Garmin Forerunners allow trackers to communicate only through
bases to which they have been previously ``paired'' or ``bonded''. Garmin's
pairing procedure works in the following manner.  The agent running on the base
searches for available ANT enable devices.  Each tracker periodically sends
broadcast beacons over the ANT interface. If the agent discovers a tracker, it
extracts its unique id (TPI). The agent uses one of two methods of
authentication: initial \textit{pairing} or \textit{passkey}.  The agent
verifies if it already stores an authfile for this TPI. If no such file exists
(i.e., this is the first time the tracker is pairing with the base), the agent
uses the \textit{pairing} method and sends a bind request to the tracker. When
prompted, the user needs to authenticate the operation, through the push of a
button on the tracker. The agent then retrieves a factory embedded ``passkey''
from the tracker. It then stores the pair $\langle TPI, passkey \rangle$ in a
newly created authfile.  During subsequent authentications, the agent uses the
\textit{passkey} method: it recovers the passkey corresponding to the TPI from
the authfile and uses it to authenticate the tracker.

The system model considered can be extended to cover the case of fitness
tracking solutions that turn the user's mobile device into a base,
e.g.,~\cite{Jawbone,Nike,Basis}.  In such systems, the agent module is a mobile
app running on the mobile device.  The tracker communicates with the smartphone
over existing network interfaces, e.g., Bluetooth or NFC. We note that Naveed
et al.~\cite{NZDWG14} identified an intriguing vulnerability of Android
smartphones bonded to health trackers.  The vulnerability stems from the fact
that the bonding occurs at smartphone device level not at the app level. This
effectively leaves the health data vulnerable to rogue apps with Bluetooth
permissions.

\noindent
{\bf The webserver.}
The online social network webserver (e.g., fitbit.com, connect.garmin.com),
allows users to create accounts from which they befriend and maintain contact
with other users. Upon purchase of a tracker and base, the user binds the
tracker to her social network account. Each social network account
has a unique id, called the \textit{user public id} (UPI). When the base
detects and sets up a connection with a nearby tracker, it automatically
collects and reports tracker stored information (step count, distance,
calories, sleep patterns) with temporal and spatial tags, to the corresponding
user's social network account. In the following, we use the term
\textit{webserver} to denote the computing resources of the online social
network.

\noindent
{\bf Tracker-to-base communication: the ANT protocol.}
Trackers communicate to bases over ANT, a 2.4 GHz bidirectional wireless
Personal Area Network (PAN) ultra-low power consumption communication
technology, optimized for transferring low-data rate, low-latency data.

\noindent
{\bf Data conversion.}
The Fitbit tracker relies on the user's walk and run stride length values to
convert the step count into the distance covered. It then extrapolates the
user's Basal Metabolic Rate (BMR)~\cite{BMR} values and uses them to convert
the user's daily activities into burned calories values. The Garmin tracker
uses the GPS receiver to compute the outdoor distance covered by the user.  It
then relies on the Firstbeat\cite{Firstbeat} algorithm to convert user data
(gender, height, weight, fitness class) and the captured heart rate information
to estimate the user's Metabolic Equivalent (MET), which in turn is used to
retrieve the calories burnt.

\subsection{Attacker Model}
\label{sec:model:attacker}

We assume that the webserver is honest, and is trusted by all participants.  We
assume adversaries that are able to launch the following types of attacks:

\noindent
{\bf Inspect attacks}. The adversary listens on the communications of trackers,
bases and the webserver.

\noindent
{\bf Inject attacks}. The adversary exploits solution
vulnerabilities to modify and inject messages into the system, as well as to
jam existing communications.

\noindent
{\bf Capture attacks}. The adversary is able to acquire trackers or bases of
victims. The adversary can subject the captured hardware to a variety of other
attacks (e.g., Inspect and Inject) but cannot access the memory of the
hardware.  We assume that in addition to captured devices, the adversary
can control any number of trackers and bases (e.g., by purchasing them).

\noindent
{\bf JTAG attacks}. JTAG and boundary scan based attacks (e.g.,~\cite{B06}),
extend the Capture attack with the ability to access the memory of captured
devices. We focus here on ``JTAG-Read'' (JTAG-R) attacks, where the attacker
reads the content of the \textit{entire} tracker memory.

\subsection{Reverse Engineering Fitbit and Garmin}
\label{sec:model:reverse}

Our goal in reverse-engineering the Fitbit Ultra and Garmin Forerunner
protocols was dual, (i) to understand the source(s) of vulnerabilities
and (ii) to develop security solutions that are interoperable with these
protocols. Sec. 103(f) of the DMCA (17 U.S.C. § 1201
(f))~\cite{Reverse.engineer} states that a person who is in legal possession of a
program, is permitted to reverse-engineer and circumvent its protection if this
is necessary in order to achieve ``interoperability''.

To log communications between trackers and webservers, we wrote USB based
filter drivers and ran them on a base. We have used Wireshark to capture all
wireless traffic between the agent software and the webserver.  To reverse
engineer Fitbit, we exploited (i) the lack of encryption in all its
communications and (ii) libfitbit~\cite{libfitbit}, a library built on
ANT-FS~\cite{ANT-FS} for accessing and transferring data from Fitbit trackers.
Unlike Fitbit, Garmin uses HTTPS with TLS v1.1 to send user login credentials.
However, similar to Fitbit, all other communications are sent over plaintext
HTTP.

Fitbit and Garmin bases both use \textit{service logs}, files that store
information concerning communications involving the base. Garmin's logs consist
of an ``authfile'' for each tracker that was paired with the base, and .FIT
files.  The authfile contains authentication information for each tracker.
Forerunner maintains 20 types of .FIT files, each storing a different type of
tracker data, including information about user activities, schedules, locations
and blood pressure readings. On the Windows installation of the Fitbit
software, daily logs are stored in cleartext in files whose names record the
hour, minute and second corresponding to the time of the first log occurrence.
Each request and response involving the tracker, base and social network is
logged and sometimes even documented in the archive folder of that log
directory.

In the following, we first focus on Fitbit's tracker memory organization and
communication protocol.

\noindent
{\bf Fitbit: Tracker memory organization.}
A tracker has both \textit{read banks}, containing data to be read by the base
and \textit{write banks}, containing data that can be written by the base. The
read banks store the daily user fitness records. The write banks store user
information specified in the ``Device Settings'' and ``Profile Settings''
fields of the user's Fitbit account. The tracker commits sensor values (step,
floor count) to the read bank once per minute. The tracker can store 7 days
worth of 1-per-minute sensor readings~\cite{FitbitSpecs}.

\begin{figure}[t]
\begin{center}
\includegraphics[width=3.1in]{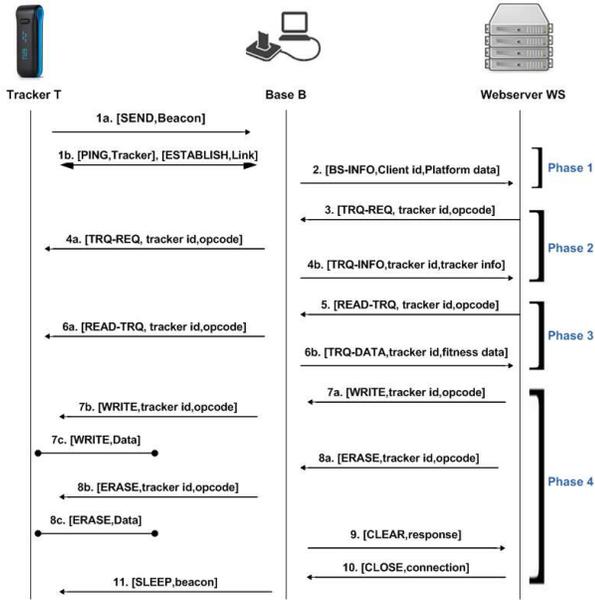}
\caption{Fitbit $Upload$ protocol. Enables the tracker to upload its collected
sensor data to the user's social networking account on the webserver. SensCrypt's
$Upload$ protocol extends this protocol, see Section~\ref{sec:solution}.}
\label{fig:fitbit:maincomm}
\end{center}
\end{figure}

The webserver communicates with the tracker through XML blocks, that contain
base64 encoded commands, or \textit{opcodes}.  tracker. Opcodes are 7 bytes
long. We briefly list below the most important opcodes and their corresponding
responses. The opcode types are also shown in Figure~\ref{fig:fitbit:maincomm}.

\begin{compactitem}

\item
{\bf Retrieve device information (TRQ-REQ):} opcode [0x24,000000].  Upon
receiving this opcode from the webserver (via the base), the tracker sends a
reply that contains its serial number (5 bytes), the hardware revision number,
and whether the tracker is plugged in on the base.

\item
{\bf Read/write tracker memory (READ-TRQ/WRITE).}
To read a memory bank, the webserver needs to issue the READ-TRQ opcode,
[0x22, $index$,00000], where $index$ denotes the memory bank requested.
The response embeds the content of the specified memory bank. To write data to
a memory bank, the webserver issues the WRITE opcode [0x23, $index$,
$datalen$,0000].  The payload data is sent along with the opcode. The
value $index$ denotes the destination memory bank and $datalen$ is the length
of the payload. A successful operation returns the response [0x41,000000].

\item
{\bf Erase memory: (ERASE)} opcode [0x25, $index$, $t$, 0].  The webserver
specifies the $index$ denoting the memory bank to be erased. The value $t$ (4
bytes, MSB) denotes the operation deadline - the date until which the data
should be erased. A successful operation returns the response [0x41,000000].

\end{compactitem}

\noindent
{\bf Fitbit: The communication protocol.}
The communication between the webserver and the tracker through the base, is
embedded in XML blocks, that contain base64 encoded opcodes: commands for the
tracker. All opcodes are 7 bytes long and vary according to the instruction
type (e.g., TRQ-REQ, READ-TRQ, WRITE, ERASE, CLEAR). The system data flow
during the data upload operation is shown in Figure~\ref{fig:fitbit:maincomm}.

\begin{enumerate}

\item
Upon receiving a beacon from the tracker, the base establishes a connection
with the tracker.

\item
{\bf Phase 1:}
The base contacts the webserver at the URL \url{HOME/device/tracker/uploadData}
and sends basic client and platform information.

\item
{\bf Phase 2:}
The webserver sends the tracker id and the opcode for retrieving tracker
information (TRQ-REQ).

\item
The base contacts the specified tracker, retrieves its information TRQ-INFO
(serial number, firmware version, etc.) and sends it to the webserver at
\url{HOME/device/tracker/dumpData/lookupTracker}.

\item
{\bf Phase 3:}
Given the tracker's serial number, the webserver retrieves the associated
tracker public id (TPI) and user public id (UPI) values. The webserver sends to
the base the TPI/UPI values along with the opcodes for retrieving fitness
data from the tracker (READ-TRQ).

\item
The base forwards the TPI and UPI values and the opcodes to the tracker,
retrieves the fitness data from the tracker (TRQ-DATA) and sends it to the
webserver at \url{HOME/device/tracker/dumpData/dumpData}.

\item
{\bf Phase 4:}
The webserver sends to the base, opcodes to WRITE updates provided by the user
in her Fitbit social network account (device and profile settings, e.g., body
and personal information, time zone, etc). The base forwards the WRITE opcode and
the updates to the tracker, which overwrites the previous values on its write
memory banks.

\item
The webserver sends opcodes to ERASE the fitness data from the tracker. The
base forwards the ERASE request to the tracker, who then erases the contents of
the corresponding read memory banks.

\item
The base forwards the response codes from the tracker to
the webserver at the address\\ \url{HOME/device/tracker/dumpData/clearDataConfigTracker}.

\item
The webserver replies to the base with the opcode to CLOSE the tracker.

\item
The base requests the tracker to SLEEP for 15 minutes, before sending its next
beacon.

\end{enumerate}

\subsection{Crypto Tools}
\label{sec:model:tools}

We use a symmetric key encryption system. We write $E_K(M)$ to denote the
encryption of a message $M$ with key $K$.  We also use cryptographic hashes
that are pre-image, second pre-image and collision resistant. We use $H(M)$ to
denote the hash of message $M$. We also use hash based message authentication
codes~\cite{HMAC}: we write $Hmac(K, M)$ to denote the authentication code
of message $M$ with key $K$.

\section{Fitbit and Garmin Attacks}
\label{sec:attacks}

During the reverse engineering process, we discovered several fundamental
vulnerabilities, which we describe here. We then detail the attacks we have
deployed to exploit these vulnerabilities, and their results.

\subsection{Vulnerabilities}
\label{sec:attacks:vulnerability}

\begin{figure}
\centering
\includegraphics[width=3.5in]{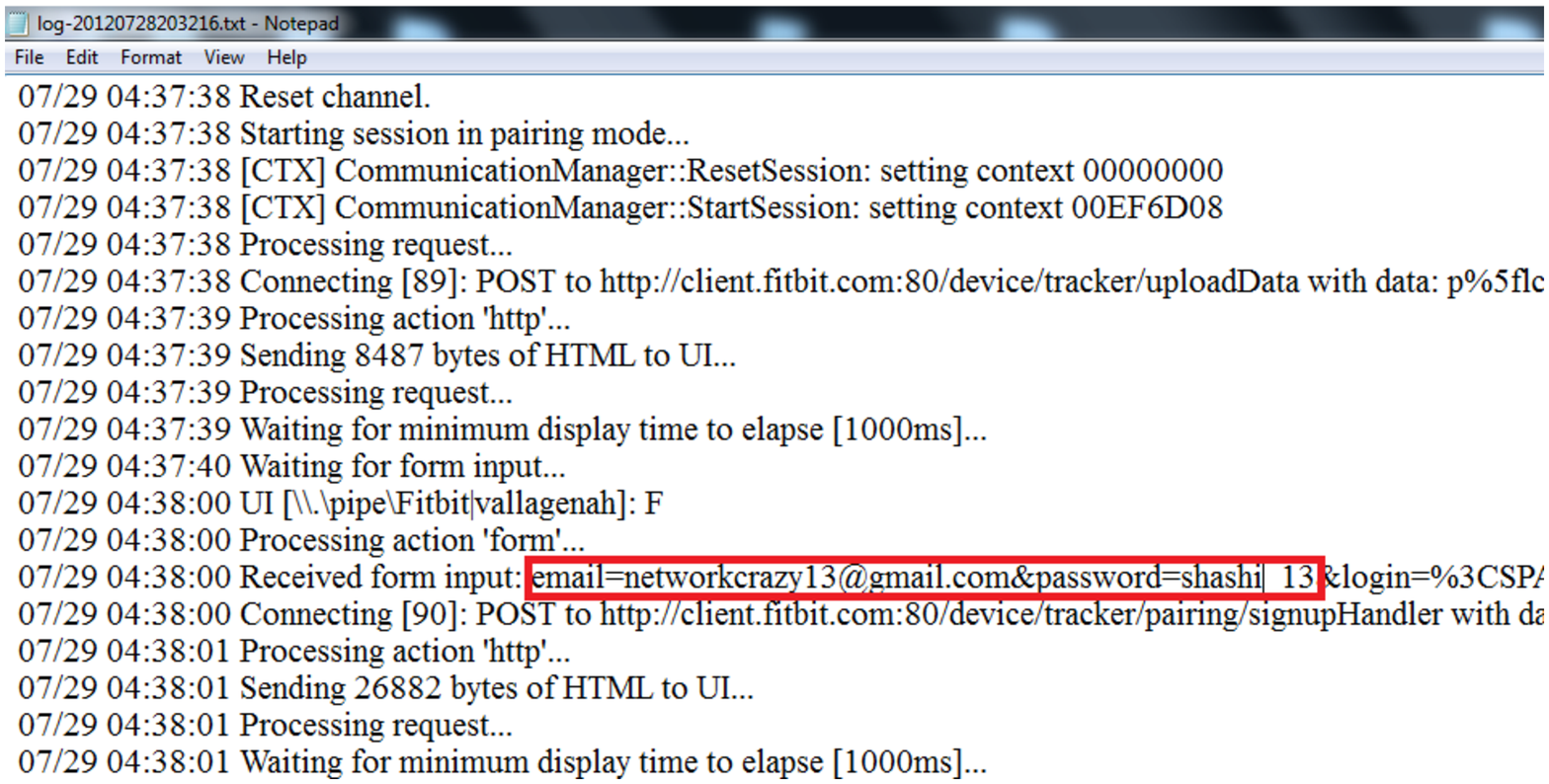}
\caption{Fitbit service logs: Proof of login credentials sent in
cleartext in a HTTP POST request sent from the base to the webserver.
\label{fig:service:log2}}
\end{figure}

\noindent
{\bf Fitbit: cleartext login information.}
During the initial user login via the Fitbit client software, user passwords
are passed to the webserver in cleartext and then stored in log files on the
base. Figure~\ref{fig:service:log2} shows a snippet of
captured data, with the cleartext authentication credentials emphasized.
Garmin uses encryption only during the login step.

\noindent
{\bf Fitbit and Garmin: cleartext HTTP data processing.}
For both Fitbit and Garmin, the tracker's data upload operation uses no
encryption or authentication. All the tracker-to-webserver communications take
place in cleartext.

\noindent
{\bf Garmin: faulty authentication during Pairing}.
The authentication in the Pairing procedure of Garmin assumes that the base
follows the protocol and has not been compromised by an attacker.  The
authentication process is not mutual: the tracker does not authenticate the
base.

\subsection{The FitBite and GarMax Tools}

We have built FitBite and GarMax, tools that exploit the above vulnerabilities
to attack Fitbit Ultra and Garmin Forerunner. FitBite and GarMax consist of
separate modules for (i) discovering and binding to a nearby tracker, (ii)
retrieving data from a nearby tracker, (iii) injecting data into a nearby
tracker and (iv) injecting data into the social networking account of a tracker
owner. We have built FitBite and GarMax over ANT-FS, in order to connect
to and issue (ANT-FS) commands to nearby trackers. The attacker needs to run
FitBite or GarMax on a base he controls.

The time required to search and bind to a nearby tracker varies significantly,
but is normally in the range of 3-20 seconds. On average, the time to query a
tracker is 12-15s. More detailed timing information is presented for the
attacks presented in the following. We conclude that these attacks can be
performed even during brief encounters with victim tracker owners.

\subsection{Attacks and Results}
\label{sec:attacks:results}

\begin{table}[b]
\centering
\begin{tabular}{l r r}
\toprule
\textbf{Type of data} & \textbf{FitBite} & \textbf{GarMax}\\
\midrule
Device info & \cmark & \cmark \\
User profile, schedules, goals & \cmark & \cmark \\
Fitness data & \cmark & \cmark \\
(GPS) Location history & \xmark & \cmark \\
\bottomrule
\end{tabular}
\caption{Types of data harvested by FitBite and GarMax from Fitbit and Garmin.
Garmin provides GPS tagged fitness information, which GarMax is able
to collect.}
\label{table:tpdc}
\end{table}

\begin{figure*}
\centering
\includegraphics[width=6.9in]{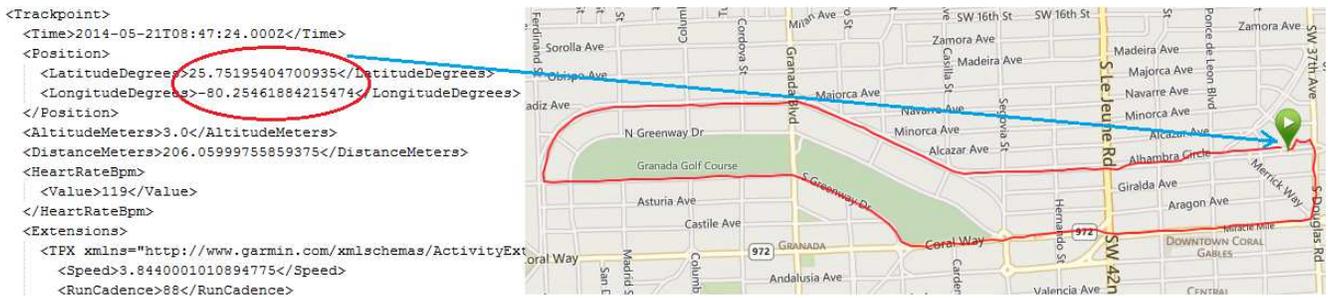}
\caption{TPDC outcome on Garmin: the attacker retrieves the user's
exercise circuit on a map (shown in red on the right side), based on individual
fitness data records (shown on the left in XML format). The data record on
the left includes both GPS coordinates, heart rate, speed and cadence.
\label{fig:garmin:tpdc}}
\end{figure*}

\noindent
{\bf Tracker Private Data Capture (TPDC).}
FitBite discovers tracker devices within transmission range and captures their
fitness information: Fitbit performs no authentication during tracker data
uploads. We exploit Garmin's assumption of an honest base to use GarMax,
running on a corrupt base, to capture data from nearby trackers.
We show how GarMax binds a ``rogue'' base agent to Garmin trackers of strangers
within a radius of 33ft. GarMax exploits the authentication vulnerability of
Garmin's Pairing procedure (see Section~\ref{sec:attacks:vulnerability}).

During the tracker authentication and passkey retrieval step of the Pairing
procedure (see Section~\ref{sec:model:reverse}), GarMax running on an attacker
controlled base, retrieves the TPI of the nearby victim tracker. It then creates a
directory with the TPI name and creates an auth file with a random, 8 byte
long passkey. GarMax verifies the tracker's serial number and other ANT
parameters, then reads the passkey from the auth file. Instead of running the
\textit{passkey} authentication method, GarMax directly downloads
fitness information (to be stored in .FIT files) from the tracker. This is
possible since the tracker assumes the base has not been corrupted, and thus
does not authenticate it.

TPDC can be launched in public spaces, particularly those frequented by
fitness users (e.g., parks, sports venues, etc) and takes less than 13s on
average.  It is particularly damaging as trackers store sensor readings (i)
with high frequency (1-9 seconds for Garmin, 1 minute for Fitbit), and (ii) for
long intervals: up to 7 days of fitness data history for Fitbit and up to 1000
laps and 100 favorite locations for Garmin. The data captured contains
sensitive user profile information and fitness information.  For Garmin this
information is tagged with GPS locations. Table~\ref{table:tpdc} summarizes the
information captured by FitBite and GarMax.

Figure~\ref{fig:garmin:tpdc} shows the reconstructed exercise circuit
of a victim, with data we recovered from a TPDC attack on Garmin.  The GPS
location history can be used to infer the user's home, locations of interest,
exercise and travel patterns.

\begin{figure}
\centering
\includegraphics[width=2.7in]{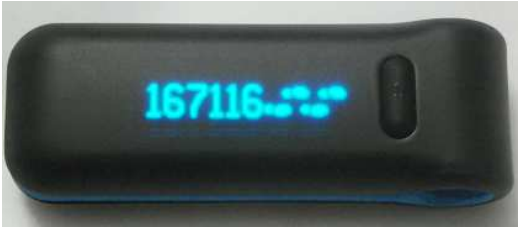}
\caption{Outcome of Tracker Injection (TI) attack on Fitbit tracker: The daily
step count is unreasonably high (167,116 steps).
\label{fig:fitbit:pic3}}
\end{figure}

\noindent
{\bf Tracker Injection (TI) Attack.}
FitBite and GarMax use the reverse engineered knowledge of the communication
packet format, opcode instructions and memory banks, to modify and inject
fitness data on neighboring trackers. On average, this attack takes less
than 18s, for both FitBite and GarMax.  Figure~\ref{fig:fitbit:pic3} shows 
a sample outcome of the TI attack on a victim Fitbit tracker, displaying
an unreasonable value for the (daily) number of steps taken by its user.

\begin{figure}
\begin{center}
\includegraphics[width=3.3in]{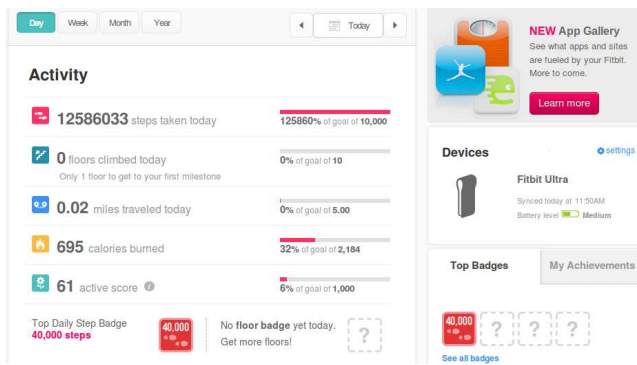}
\caption{Snapshot of Fitbit user account data injection attack. In addition
to earning undeserved badges (e.g., the ``Top Daily Step''), it enables insiders
to accumulate points and receive financial rewards through sites
like Earndit~\cite{Earndit}.}
\label{fig:attack:inject}
\end{center}
\end{figure}

\noindent
{\bf User Account Injection (UAI) Attack.}
We used FitBite and GarMax to report fabricated fitness information into our
social networking accounts. We have successfully injected unreasonable daily
step counts, e.g., 12.58 million in Fitbit, see Figure~\ref{fig:attack:inject}.
Fitbit did not report any inconsistency, especially as the corresponding
distance we reported was 0.02 miles! The UAI attack takes only 6s on average.

Similarly, GarMax fabricates an activity file embedding the attacker provided
fitness data in FIT/TCX~\cite{TCX} format.  The simplest approach is to copy an
existing activity file of the same or another user (made publicly available in
the Garmin Connect website) and modify device and user specific information.
We have used GarMax to successfully inject ``running'' activities of 1000 miles
each, the largest permissible value, while keeping the other parameters
intact.

\noindent
{\bf Free Badges and Financial Rewards.}
By successful injection of large values in their social networking accounts,
FitBite and GarMax enable insiders to achieve special milestones and acquire
merit badges, without doing the required work. Figure~\ref{fig:attack:inject}
shows how in Fitbit, the injected value of 12.58 million steps, being greater
than 40,000, enables the account owner to acquire a ``Top Daily Step'' badge.
Furthermore, by injecting fraudulent fitness information into
Earndit~\cite{Earndit}, an associated site, we were able to accumulate
undeserved rewards, including 200 Earndit points, redeemable for a \$20 gift
card.

\begin{figure}
\begin{center}
\includegraphics[width=3.1in]{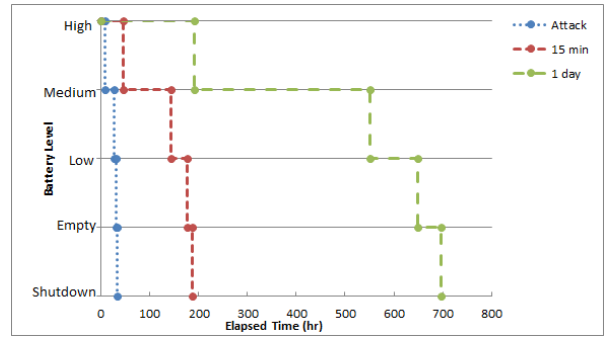}
\caption{Battery drain for three operation modes. The \textit{attack} mode
drains the battery around 21 times faster than the 1 day upload mode and 5.63
times faster than the 15 mins upload mode.}
\label{fig:fitbit:battery}
\end{center}
\end{figure}

\noindent
{\bf Battery Drain Attack.}
FitBite allows the attacker to continuously query trackers in her vicinity,
thus drain their batteries at a faster rate. To understand the efficiency of
this attack, we have experimented with 3 operation modes. First, the
\textit{daily upload} mode, where the tracker syncs with the USB base and the
Fitbit account once per day. Second, the \textit{15 mins upload} mode, where
the tracker is kept within 15 ft. from the base, thus allowing it to be queried
once every 15 minutes. Finally, the \textit{attack} mode, where FitBite's TM
module continuously (an average of 4 times a minute) queries the victim
tracker. To avoid detection, the BM module uploads tracker data into the
webserver only once every 15 minutes. Figure~\ref{fig:fitbit:battery} shows
our battery experiment results for the three modes: FitBite drains the tracker
battery around 21 times faster than the 1 day upload mode and 5.63 times
faster than the 15 mins upload mode.

In the daily upload mode, the battery lasted for 29 days.  In the 15 mins
upload mode, the battery lasted for 186.38 hours (7 days and 18 hours).  In
the attack mode, the battery lasted for a total of 32.71 hours. While this
attack is not fast enough to impact trackers targeted by casual attackers, it
shows that FitBite drains the tracker battery around 21 times faster than the
1 day upload mode and 5.63 times faster than the 15 mins upload mode.



\noindent
{\bf Denial of Service.}
FitBite's injection attack can be used to prevent Fitbit users from correctly
updating their real-time statistics. The storage capacity of the Garmin
tracker is limited to 1000 laps. Thus, an attacker able to injects a number of
fake laps exceeding the 1000 limit, can prevent the tracker from recording the
user's valid data.  A Fitbit tracker can display up to 6 digit values.
When the injected value exceeds 6 digits, the least significant digits
can not be displayed on the tracker. This prevents the user from keeping track
of her daily performance evolution.  In addition, for both Fitbit and Garmin,
the attacker can render part of the recorded data useless, by injecting
incorrect user profile information. For instance, by modifying user profile
information (e.g., height, weight, see Section~\ref{sec:model:system}), the
attacker corrupts information built based on it, e.g., ``calories burnt''.

\section{A Protocol for Lightweight Security}
\label{sec:solution}

\subsection{Solution Requirements}
\label{sec:solution:requirements}

We aim to develop a solution for low power fitness trackers that satisfies the
following requirements:

\begin{compactenum}

\item
{\bf Security.}
Defend against the attacks described in Section~\ref{sec:model:attacker}.

\item
{\bf Minimal tracker overhead.} Minimize the computation and storage overheads
imposed on the resource constrained trackers.

\item
{\bf Flexible upload.}
Allow trackers to securely upload sensor information through multiple bases.

\item
{\bf User friendly.} Minimize user interaction.

\item
{\bf Level tracker memory wear.} Extend memory lifetime by leveling the
wear of its blocks.

\end{compactenum}

\subsection{Public Key Cryptography: A No Go}
\label{sec:solution:pkc}

We propose first FitCrypt, a solution to explore the feasibility of
public key cryptosystems to efficiently secure the storage and communications
of trackers. In FitCrypt, each tracker stores a public key. The corresponding
private key is only known by the webserver. Each sensor data record is
encrypted with the public key before being stored on the tracker.  RSA with a
2048 bit key imposes a 4-hold storage overhead on Fitbit (each record of 64B is
converted into a 256B record) and a 3.2-hold overhead on Garmin.  We also
consider ECIES (Elliptic Curve Integrated Encryption Scheme), an elliptic curve
crypto (ECC) solution that uses a 224 bit key size, the security equivalent of
RSA with 2048 bit modulus. ECIES imposes a storage overhead of $224+3r$ bits,
where $r=112$ is the size of a security parameter.  Thus, the storage overhead
is 165\% for Fitbit and 150\% for Garmin).

When run on an Arduino Uno board, FitCrypt-RSA takes 2.3s and FitCrypt-ECC
takes 2.5s to encode a single sensor record (see Table~\ref{table:record:data},
Section~\ref{sec:evaluation}).  Garmin records sensor data with a frequency as
high as one write per second. FitCrypt imposes a 250\% overhead on the sensor
recording task (of 2.5s every 1s interval), thus does not satisfy the second
requirement of Section~\ref{sec:solution:requirements}. To address this issue,
in the following we introduce SensCrypt, a lightweight and secure solution for
wearable trackers.

\subsection{SensCrypt}

We introduce SensCrypt, a lightweight protocol for providing secure data
storage and communication in fitness centric social sensor networks.

\noindent
{\bf Protocol overview.}
Let $U$ denote a user, $T$ denote her tracker, $B$ a base and $W$ the
webserver. $T$'s memory is divided into records, each storing one
snapshot of sensor data. The memory is organized using a circular buffer
structure, to ensure an even wear. $T$ shares a symmetric key $K_T$ with $W$.
$W$ also maintains a unique {\emph secret} key $K_W$ for each tracker $T$.

To prevent Inject attacks, all communications between $T$ and $W$ are
authenticated with $K_T$.  To prevent Inspect, Capture and JTAG-R attacks, we
encode each tracker record using two pseudo-random numbers (PRNs).  One PRN is
generated by $W$ using $K_W$ and written on $T$ during data sync protocols. The
other PRN is generated by $T$ using $K_T$ at the time when the record is
written on its memory. Both PRNs can later be reconstructed by $W$. This
approach significantly increases the complexity of an attack: the attacker
needs to capture the encoded data and both PRNs to recover the cleartext data.

\subsection{The SensCrypt Protocol}
\label{sec:solution:details}

\begin{table}
\centering
\begin{tabular}{l r}
\toprule
\textbf{Notation} & \textbf{Definition}\\
\midrule
$U$, $T$, $B$, $W$ & user, tracker, base and webserver\\
$id_U$, $id_T$, $id_B$  & unique identifiers of $U$, $T$ and $B$ \\
\dirty & pointer to first written record\\
\clean & pointer to first available record\\
\start, \textit{end} & pointers to memory bounds\\
$K_W$ & symmetric key maintained by $W$ for $T$\\
$K_T$ & symmetric key shared by $W$ and $T$\\
$ctr$ & counter shared by $W$ and $T$\\
$Map$ & data base of $W$ for users and trackers\\
$mem$ & memory of a tracker\\
\bottomrule
\end{tabular}
\caption{Symbol definitions.}
\label{table:notation}
\end{table}

\begin{figure}
\centering
\includegraphics[width=3.1in]{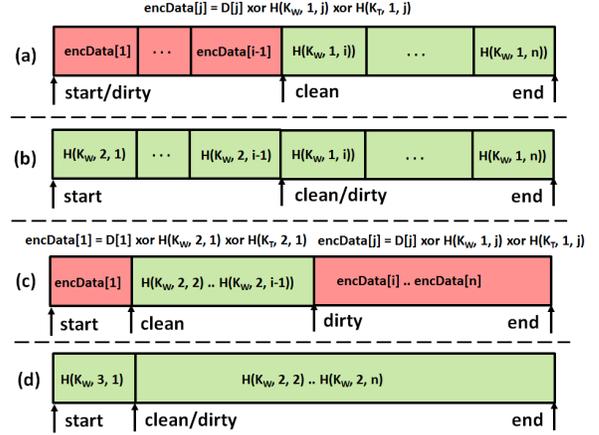}
\vspace{-5pt}
\caption{Example SensCrypt tracker memory ($mem$). Light green denotes ``clean'',
unwritten areas.  Red denotes areas that encode tracker sensor data.
(a) After ($i$-1) records have been written. The $ctr$ is 1.
(b) After $Upload$ occurs at the state in (a). The $ctr$ becomes 2, to enable
the creation of fresh PRNs, overwritten on the former red area.
(c) After $n-i+2$ more records have been written from state (b), leading to
the \clean\ pointer cycling over from the start of the memory.
(d) After $Upload$ occurs at the state in (c).
\label{fig:sensio:memory}}
\end{figure}

Let $id_U$, $id_B$, and $id_T$  denote the public unique identities of $U$,
$B$, and $T$. $U$ has an account with $W$. $W$ manages a database $Map$ that
has an entry for each user and tracker pair: $Map[id_U,id_T]$ = $[id_U$,
$id_T$, $K_T$, $K_W$, $ctr]$. Each tracker is factory initialized with a
symmetric key $K_T$ and a counter $ctr$ initialized to 1. $K_T$ and $ctr$ are
also stored in $Map[id_U,id_T]$. $K_W$ is a per-tracker symmetric key, kept
secret by $W$. Table~\ref{table:notation} summarizes these symbols for easy
access.

SensCrypt consists of 2 procedures, $RecordData$ and $Upload$. $RecordData$ is
invoked by $T$ to record new sensor data; $Upload$ allows it to sync
its data with $W$. We now describe the organization of the tracker
memory.

\noindent
{\bf Tracker Memory Organization}.
Let $mem$ denote the memory of $T$. $mem$ is divided in ``records'' of fixed
length (e.g., 64 bytes for Fitbit, 80 bytes for Garmin). Each record stores one
report from the tracker's sensors (see Section~\ref{sec:model:system}).  We
organize time into fixed length ``epochs'' (e.g., 2s long for Fitbit, 1-9s long
for Garmin). $RecordData$ records sensor data once per epoch. $mem$ is
organized using a circular buffer. The \dirty\ pointer is to the location of
the first written record, and the \clean\ pointer is to the location of the
first record available for writing. When reaching the end of $mem$, both
records ``circle'' over to the \start\ pointer.
Figure~\ref{fig:sensio:memory} illustrates the SensCrypt tracker storage
organization, after the execution of various $RecordData$ and $Upload$
procedures. Algorithm~\ref{alg:senscrypt} shows the pseudo-code of the
procedures.

During $Upload$, each previously written tracker record is reset by $W$ to
store a pseudo-random value (line 18 and lines 21-29 of
Algorithm~\ref{alg:senscrypt}). That is, the $i$-th record of the tracker's
memory is set to hold $E_{K_W}(ctr, i)$, where $K_W$ is the secret key $W$
stores for $T$. The index $i$ ensures that each record contains a different
value. $ctr$ counts the number of times $mem$ has been completely overwritten;
it ensures that a memory record is overwritten with a different encrypted
value.

\renewcommand{\baselinestretch}{1.0}
\begin{algorithm}[t]
\begin{minipage}{0.45\textwidth}
\begin{tabbing}
XX\=XX\=XX\=XX\=XX\=X\= \kill

1.$\mathtt{\mbox{\bf{Object}}\ implementation\ Memory;}$\\
2.\>$\mathtt{T: mem:\ record[];\qquad\ \quad \quad \quad \ \ \# tracker\ memory}$\\
3.\>$\mathtt{T: dirty:\ int;\ \# pointer\ to\ used\ area}$\\
4.\>$\mathtt{T: clean:\ int;\ \# pointer\ to\ unused\ area}$\\
5.\>$\mathtt{T: start, end:\ int;\ \# memory\ bounds}$\\
6.\>$\mathtt{W: K_W:\ byte[];\ \# key\ for\ T}$\\
7.\>$\mathtt{W,T: K_T:\ byte[];\ \# key\ shared\ by\ T, W}$\\
8.\>$\mathtt{W,T: ctr:\ int;\ \# counter\ shared\ by\ T, W}$\\\\

9.\>$\mathtt{\mbox{\bf{Operation}}\ int\ \underline{T: RecordData}(D: sensor\ data)}$\\
10.\>\>$\mathtt{mem[clean]\ \oplus = D \oplus E_{K_T}(ctr, clean);}$\\
11.\>\>$\mathtt{clean = clean + 1;}$\\
12.\>\>$\mathtt{\mbox{\bf{if}}\ (clean == end)\ \mbox{\bf{then}};}$\\
13.\>\>\>$\mathtt{clean = start;\ \mbox{\bf{fi}}}$\\
14.\>$\mathtt{\mbox{\bf{end}}}$\\

15.\>$\mathtt{\mbox{\bf{Operation}}\ void\ \underline{ProcessRecord}(ind: int, c: int)}$\\
16.\>\>$\mathtt{W:\ D = mem[ind] \oplus E_{K_W}(c, ind) \oplus E_{K_T}(c, ind);}$\\
17.\>\>$\mathtt{W:\ process(D);}$\\
18.\>\>$\mathtt{W \rightarrow T:\ mem[ind] = E_{K_W}(c+1, ind);}$\\
19.\>$\mathtt{\mbox{\bf{end}}}$\\

20.\>$\mathtt{\mbox{\bf{Operation}}\ void\ \underline{Upload}()}$\\
21.\>\>$\mathtt{\mbox{\bf{if}}\ (dirty < clean)\ \mbox{\bf{do}}}$\\
22.\>\>\>$\mathtt{\mbox{\bf{for}}\ (i=dirty; i < clean; i++)\ \mbox{\bf{do}}}$\\
23.\>\>\>\>$\mathtt{ProcessRecord(i, ctr);\ \mbox{\bf{od}}}$\\

24.\>\>$\mathtt{\mbox{\bf{else\ if}}\ (clean < dirty)\ \mbox{\bf{do}}}$\\
25.\>\>\>$\mathtt{\mbox{\bf{for}}\ (i=dirty; i \le end; i++)\ \mbox{\bf{do}}}$\\
26.\>\>\>\>$\mathtt{ProcessRecord(i, ctr);\ \mbox{\bf{od}}}$\\
27.\>\>\>$\mathtt{\mbox{\bf{for}}\ (i=start; i < clean; i++)\ \mbox{\bf{do}}}$\\
28.\>\>\>\>$\mathtt{ProcessRecord(i, ctr+1);\ \mbox{\bf{od}}}$\\
29.\>\>\>$\mathtt{W,T:\ ctr = ctr+1;\ \mbox{\bf{fi}}}$\\
30.\>\>$\mathtt{T:\ dirty = clean;}$\\
31.\>$\mathtt{\mbox{\bf{end}}}$\\
\end{tabbing}
\caption{Tracker memory management pseudocode. Instructions
preceded by $W:$ are executed at the webserver, those preceded by $T:$ are
executed at the tracker. $W \rightarrow T: I$ denotes an instruction $I$ issued
at $W$ and executed at $T$. The entire RecordData is executed at $T$.
Figure~\ref{fig:sensio:memory} illustrates the pseudocode.}
\label{alg:senscrypt}
\end{minipage}
\normalsize
\vspace{-10pt}
\end{algorithm}

\noindent
{\bf The RecordData Procedure}.
Commit newly recorded sensor data $D$ to $mem$, in the next available record,
pointed to by \clean. $T$ generates a new pseudo-random value, $E_{K_T}(ctr,
\clean)$, and xors it into place with $mem[\clean]=E_{K_W}(ctr, clean)$ and
$D$ (see line 10 of algorithm~\ref{alg:senscrypt}):
\[
mem[\clean] = D \oplus E_{K_T}(ctr, \clean) \oplus E_{K_W}(ctr, clean).
\]
The \clean\ pointer is then incremented (line 11).  When reaching the
\textit{end} of $mem$, \clean\ circles back to \start (lines 12,13). We call
``red'' the written records and ``green'' the records available for write.
\dirty\ and \clean\ enable us to reduce the communication overhead of $Upload$
(see next): instead of sending the entire $mem$, $T$ sends to $W$ only the red
records.

\noindent
{\bf The Upload Procedure}.
We present the SensCrypt $Upload$ as an extension of the corresponding Fitbit protocol
illustrated in Figure~\ref{fig:fitbit:maincomm}.  In the following, each
message $M$ sent between $T$ and $W$ is accompanied by an authentication value
$Hmac(K_T, M)$, where Hmac is a hash based message authentication
code~\cite{HMAC}. The receiver of the message uses $K_T$ to verify the
authenticity of the sender and of the message. For simplicity of exposition, in
the following we omit the Hmac value.

$Upload$ extends steps 6b and 7 of the Fitbit $Upload$. Specifically, when $T$
receives the READ-TRQ command (step 6a), it compares the \dirty\ and \clean\
pointers.  If $\dirty < \clean$ (see Figure~\ref{fig:sensio:memory}(a)), $T$
sends to $W$, through $B$,
\[
{\tt T \rightarrow B \rightarrow W: TRQ-DATA, id_T, mem[dirty .. clean]},
\]
where $mem[dirty .. clean]$ denotes $T$'s red memory area. For each record $i$
between \dirty\ and \clean, $W$ uses keys $K_T$ and $K_W$ and the
current value of $ctr$ to recover the sensor data: $D[i] = mem[i] \oplus
E_{K_T}(ctr, i) \oplus E_{K_W}(ctr, i)$ (see lines 21-23 and line 16). Then, in step 7 of Upload (see
Figure~\ref{fig:fitbit:maincomm}), $W$ sends to $T$:
\[
{\tt W \rightarrow B \rightarrow T: WRITE, id_T, E_{K_T}(ctr+1, E_{K_W}(ctr+1, i))},
\]
$\forall i$=\dirty..\clean. $T$ uses $K_T$ to decrypt each $E_{K_T}(ctr+1,
E_{K_W}(ctr+1, i))$ value. If the first field of the result equals $ctr+1$, $T$
overwrites $mem[\dirty+i]$ with $E_{K_W}(ctr+1, i)$ (see line 18), then sets
\dirty=\clean (line 30).  Thus, mem[\dirty .. \clean] becomes green.
The case
where \clean\ $<$ \dirty, occurring when \clean\ circles over, past the memory
end, is handled similarly, see lines 24-29 of Algorithm~\ref{alg:senscrypt} and
Figure~\ref{fig:sensio:memory}(c) and (d). We eliminate the ERASE communication
(steps 8 and 9 in Figure~\ref{fig:fitbit:maincomm}) from the Fitbit protocol.

\section{Analysis}
\label{sec:analysis}

\subsection{SensCrypt Advantages}
\label{sec:analysis:advantages}

SensCrypt ensures an even wear of tracker memory: the most overwritten memory
record has at most 2 overwrites more than the least overwritten record.  To see
why this is the case, consider that once written, a record is
not overwritten until a next $Upload$ takes place. The circular buffer organization
of the memory ensures that all the memory records of the tracker are
overwritten, not just the ones at the start of the memory. Using the example
illustrated in Figure~\ref{fig:sensio:memory}(d), notice that the first
record, has been overwritten twice since the subsequent green blocks: once with
$encData[1]$, see Figure~\ref{fig:sensio:memory}(c), and once with the new
$E_{K_W}(3, 1)$ received from $W$.

By preventing excessive overwriting of records at the beginning of the memory,
SensCrypt extends the life of trackers. This is particularly important for
flash memories, that have a limited number of P/E (program/erase) cycles.

SensCrypt is user friendly, as the user is not involved in $Upload$ and
$RecordData$ procedures. The SensCrypt base is thin, required to only setup
standard secure SSL connections to $W$, and forward traffic between $T$ and
$W$. SensCrypt imposes no storage overhead on trackers: sensor data is xor-ed
in-place in $mem$.

\subsection{Security Discussion}
\label{sec:analysis:security}

\begin{table}
\centering
\begin{tabular}{l r r}
\toprule
\textbf{Capabilities} & \textbf{SensCrypt} & \textbf{FitCrypt}\\
\midrule
Inspect & TPDC, TI, UAI & TPDC, TI, UAI\\
Inject & TPDC, TI, UAI & TPDC, TI, UAI\\
Capture & TPDC, TI, UAI & TPDC, TI, UAI\\
JTAG-R & TPDC, TI, UAI & TPDC, TI, UAI\\
JTAG-RW & TPDC & TPDC\\
JTAG-R + Inspect & TI, UAI & TPDC, TI, UAI\\
JTAG-R + Inject & TI & TPDC, TI\\
Double JTAG-R & TI, UAI & TPDC, TI, UAI\\
\bottomrule
\end{tabular}
\caption{
Comparison of defenses provided by SensCrypt and FitCrypt against the
types of attacks described in Section~\ref{sec:attacks:results} when the
adversary has a combination of the capabilities described in
Section~\ref{sec:model:attacker}.  Each element in the table describes which
attacks are thwarted by the corresponding solution.}
\label{table:analysis}
\end{table}

Consider the life cycle of record $i$, $R_i$, on $T$. After the execution of
the first $Upload$, $R_i$ is initialized with $E_{K_W}(ctr, i)$.  When
$R_i$ is overwritten with sensor data, it contains $encData[i]$ = $D[i] \oplus
E_{K_T}(ctr, i) \oplus E_{K_W}(ctr, i)$. Subsequently, $R_i$ is not touched
until an Upload takes place. During Upload, the (encoded) content of $mem[i]$
is sent to $W$, who subsequently overwrites $R_i$ with a new value:
$E_{K_W}(ctr+1, i)$.

The base does not contribute to the messages it
forwards between $T$ and $W$. Hence, the base does not need to be
authenticated. The use of the $ctr+1$ value in communications through the base
ensures message freshness.

Without $E_{K_T}(ctr, i)$, an Inspect adversary capturing communications
between $T$ and $W$ cannot recover $mem[i]$. The use of HMACs with the key
$K_T$ to authenticate communications between $T$ and $W$ prevents Inject
attacks: an attacker that modifies existing messages or injects new messages
cannot create valid HMAC values.

An attacker that launches a Capture attack against a victim tracker or base,
cannot recover information from them and thus has no advantage over general
Inspect and Inject attacks.  An adversary that captures a tracker $T$ and
launches a JTAG-R attack can either read $E_{K_W}(ctr, i)$ or $D[i]$ $\oplus$
$E_{K_W}(ctr, i)$ $\oplus$ $E_{K_T}(ctr, i)$, but not both. The use of the
$E_{K_T}(ctr, i)$ value prevents an attacker from recovering $D[i]$. A JTAG-R
attack against a captured, trusted base of tracker $T$ offers no advantage over
Inspect and Inject attacks: in SensCrypt, the base only forwards traffic
between $T$ and $W$. Similar to JTAG-R, a JTAG-RW attack against a captured
tracker cannot decode previously encoded sensor data; it can however encode
fraudulent data on the tracker (TI attack) and thus also inject data into $W$
(UAI attack).

An adversary able to perform Inspect, Capture and JTAG-R attacks can gain
access to $E_{K_W}(ctr, i)$ when sent by $W$, then use JTAG-R to read $T$'s
$K_T$, compute $E_{K_T}(ctr, i)$ and learn $D[i]$ (TPDC attack).  We note the
complexity of this attack. If able to further implement Inject attacks, the
adversary can also succeed in a UAI attack.

Furthermore, SensCrypt is vulnerable to an adversary able to capture $T$ twice,
at times $t_1$ and $t_2$.  At time $t_1$ the adversary uses JTAG-R to read
$E_{K_W}(ctr, i)$. At time $t_2$, assuming $T$ has already written record $i$,
the adversary uses JTAG-R to read $mem[i]$ and $K_T$ and recover $D[i]$. This
$\textit{double JTAG-R attack}$ is significantly more complex than a single
JTAG-R attack. In addition, this attack is further complicated by time
constraints: At $t_1$, record $i$ has not yet been written, and at $t_2$ it has
been written \textit{but} an Upload has not yet been executed. An Upload
procedure before $t_2$ would overwrite record $i$ with $E_{K_W}(ctr+1, i)$,
effectively thwarting this attack.

FitCrypt is resilient to TPDC attacks launched by adversaries capable of
performing JTAG-R and Inspect, Inject and double JTAG-R attacks: $T$'s records
encrypted with the public key can only be decrypted by $W$.
Table~\ref{table:analysis} summarizes the comparison of SensCrypt and FitLock
defenses. While providing more defenses (i.e., against TPDC
for several attacker capabilities), FitLock is not a viable solution on
most of the available trackers (see Section~\ref{sec:evaluation}).

\section{Applications}

\begin{table*}
\centering
\begin{tabular}{l r r r r}
\toprule
\textbf{Platform} & \textbf{Type of data} & \textbf{Communication} & \textbf{Coverage} & {Memory}\\
\midrule
Fitbit~\cite{Fitbit} & user profile, fitness, sleep data & ANT+, BT & 5-50m & 96 KB RAM, 112 KB flash\\
Garmin FR610~\cite{Forerunner} & fitness data, heart rate, location & ANT+ & 10-20m & 1 MB\\
Nike+~\cite{Nike} & profile, fitness data & BT & 50m & Flash 256KB, RAM 32 KB\\ 
Jawbone Up~\cite{Jawbone} & fitness, sleep data & BT & 50m & 128KB Flash, 8KB RAM\\ 
Motorola MotoActv~\cite{MotoActv}& fitness data, user profile & ANT+, BT, Wi-Fi & 35m & 16 GB\\
Basis B1~\cite{Basis} & fitness, sleep data, heart rate & BT & 50m & 7 days of data\\
Mother~\cite{Mother} & motion, fitness, proximity & 915-MHz & 30m & 32 KB RAM\\
Nest~\cite{Nest} & utility data & Wi-Fi & 35m & 512Mb DRAM, 2 Gb flash\\ 
Belkin WeMo~\cite{WeMo} & home electronics & Wi-Fi & 35m & RAM 32 MB, Flash 16 MB\\
\bottomrule
\end{tabular}
\caption{SensCrypt applicability: fitness trackers, home monitoring solutions.}
\label{table:application}
\end{table*}

SensCrypt can be applied to a range of sensor based platforms, where resource
constrained sensors are unable to directly sync their data with a central
webserver and need to use an Internet connected base. This includes a large
number of popular fitness and home monitoring solutions.
Table~\ref{table:application} summarizes several such platforms, including the
communication and storage capabilities of their sensors.

SensCrypt can also be used in applications where the sensors need to be
immobile, while being able to operate without network connectivity.  Examples
include health, infrastructure, traffic, building and campus monitoring
solutions. The bases through which the sensors sync with the webserver are
mobile, e.g., smartphones of workers, who may become proximal to the sensors
with the intention of data collection or as a byproduct of routine operations.

SensCrypt can also secure the data and communications of platforms for social
psychological studies. One such example is SociableSense~\cite{RMMR11}, a
smartphone solution that captures sensitive user behaviors (including
co-location), processes the information on a remote server, and provides
measures of user sociability.

\section{Sens.io: The Platform}
\label{sec:implementation}

\begin{figure}
\begin{center}
\includegraphics[width=2.9in]{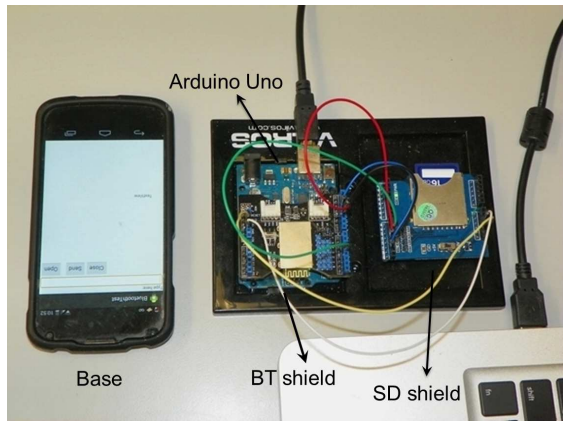}
\caption{Testbed for SensCrypt. Sens.io is the Arduino Uno device equipped
with Bluetooth shield and SD card is the tracker. Nexus 4 is the base.}
\label{fig:sens.io}
\end{center}
\end{figure}

We have built Sens.io, a prototype tracker, from off-the-shelves components. It
consists of an Arduino Uno Rev3 ~\cite{arduinoUno} and external Bluetooth
(Seeeduino V3.0) and SanDisk card shields. The Arduino platform is a good model
of resource constrained trackers: its ATmega328 micro-controller has a 16MHz
clock, 32 KB Flash memory, 2 KB SRAM and 1KB EEPROM.
The Bluetooth card
has a default baud rate of 38,400 and communication range up to 10m.
Since the Arduino has 2 KB SRAM, it can only rely on 1822 bytes to buffer data
for transmissions. The SD card (FAT 16) can be accessed at the granularity of
512 byte blocks.

The cost of Sens.io is \$52 (\$25 Arduino card, \$20 Bluetooth shield, \$2.5 SD
Card shield, \$4 SD card, see Figure~\ref{fig:sens.io}), a fraction of Fitbit's
(\$99) and Garmin's (\$299) trackers.

\begin{figure}
\begin{center}
\includegraphics[width=3.3in]{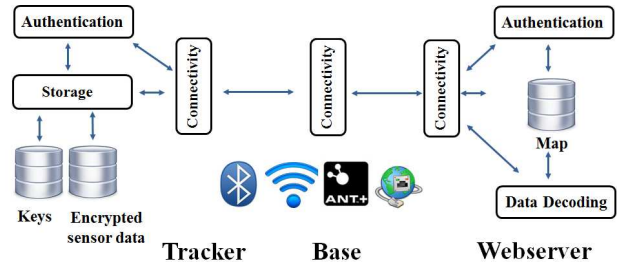}
\caption{SensCrypt architecture.  The tracker relies on locally stored key
$K_T$ to authenticate webserver messages and encode sensor data.  The webserver
manages the $Map$ structure, to authenticate and decrypt tracker reports.}
\label{fig:senscrypt:solution}
\end{center}
\end{figure}

\noindent
{\bf SensCrypt.}
We have implemented a general, end-to-end SensCrypt architecture, as
illustrated in Figure~\ref{fig:senscrypt:solution}. We have implemented the
tracker both in Arduino's programming language (a Wiring
implementation~\cite{ArduinoGuide}), and, for generality, in Android. The base
component (written exclusively in Android) is a simple communication relay.  We
implemented the webserver using Apache Tomcat 7.0.52 and Apache Axis2 Web
services engine. We used the MongoDB 2.4.9 database to store the $Map$
structure.  We implemented a Bluetooth~\cite{Bluetooth} serial communication
protocol between the tracker and the base.

\noindent
{\bf The testbed.} We used Sens.io for the tracker, an Android Nexus 4 with
1.512 GHz CPU for the base, and a 2.4GHz Intel Core i5 Dell laptop with 4GB of
RAM for the webserver.  We used Bluetooth for tracker to base communications
and Wi-Fi for the connectivity between the base and the webserver.
Figure~\ref{fig:sens.io} illustrates our testbed.


\section{Evaluation}
\label{sec:evaluation}

We used Sens.io for the tracker, Android Nexus 4 with 1.512 GHz CPU for the
base, and a 2.4GHz Intel Core i5 Dell laptop with 4GB of RAM for the webserver.
We used Bluetooth for tracker to base communications and Wi-Fi for the
connectivity between the base and the webserver.  Figure~\ref{fig:sens.io}
illustrates our testbed.

In the following, we report evaluation results, as averages taken over at least
10 independent protocol runs.

\begin{table}
\centering
\begin{tabular}{l r r r}
\toprule
\textbf{Platform} & \textbf{SensCrypt} & \textbf{FitCr-RSA} & \textbf{FitCr-ECC}\\
\midrule
Fitbit & 6.02 & 2300 & 2520\\
Garmin & 6.06 & 2300 & 2520\\
\bottomrule
\end{tabular}
\caption{RecordData: computation overhead in ms. FitCrypt-RSA 2048
bit is not viable on Arduino (2.3s). FitCrypt-ECC 224 bit (equivalent of RSA
2048 bit) is even less efficient. SensCrypt is 2-3 orders of magnitude more
efficient.}
\label{table:record:data}
\end{table}

\subsection{Tracker: RecordData Overhead}

We have investigated the overhead of the $RecordData$ procedure on Sens.io.
Table~\ref{table:record:data} compares the performance of SensCrypt and
FitCrypt, with times shown in milliseconds. We have explored two versions of
FitCrypt, using RSA and ECC. FitCrypt-RSA with a 1024 bit modulus takes more
than 500ms, but is currently obsolete. FitCrypt-RSA with a 2048 bit modulus
hangs on Sens.io due to its low (2KB) RAM. The value shown in
Table~\ref{table:record:data} is from~\cite{GPWES04}, where a similar platform
was used.  FitCrypt-ECC uses ECIES, an elliptic curve cryptography solution,
with a 224 bit key size, the security equivalent of RSA with 2048 bit modulus.
FitCrypt-RSA 2048 and FitCrypt-ECC are not viable alternatives, imposing an
overhead of 230\% for 1 per sec. RecordData frequency. SensCrypt imposes
however an overhead of less than 1\% (6ms for each 1s interval between
RecordData runs).

\subsection{Webserver: Storage Overhead}

The webserver maintains a data structure, $Map$, with a record for each user
and tracker pair. The entry consists of user, tracker and bases ids (8 byte
long each), a salt (16B), password hash (28B), 2 symmetric keys (32B each) and
a counter (1B). Assuming a single base in the $Bases$ list, a Map entry stores
133 bytes. For a 1 million user base, the webserver needs to store a $Map$
structure of 127MB. The average time to retrieve a record from a 1 million user
$Map$ is 158ms.

\subsection{Upload: End-to-end Overhead}

We consider a ``Fitbit'' scenario where the Upload procedure runs once every 15
minutes when in the vicinity of a base. Assuming a RecordData frequency of once
every 2s (usual in Garmin), and a record
size of 64B, SensCrypt uploads and overwrites 71 blocks of 512B each. The
tracker side of the SensCrypt Upload procedure takes 502ms, dominated by the
cost to read and write 71 blocks of data from/to the SD card.  A single core of
the Dell laptop can support 5 Uploads per second. The server cost is dominated
by the 158ms cost of retrieving a record from a 1 million entry $Map$. The
Upload/s rate of the webserver can be improved by caching the least recently
accessed or most popular records of $Map$.  Even though transferring over
Bluetooth, the communication cost of SensCrypt's Upload is 153ms. This is due
to the low RAM available on Arduino for buffering packets (2KB).

SensCrypt's total Upload time of 845ms is 400ms less than FitCrypt's, assuming
Fitbit's memory size. We note however that Fitbit records data only once per
minute, a rate at which SensCrypt would perform significantly faster.
SensCrypt is 13 times faster (by more than 11s) than FitCrypt when considering
Garmin's memory (2000 blocks of 512B). This gain is due to SensCrypt's
optimization of only uploading the red, written blocks, instead of the entire
memory.

Furthermore, even on the communication restricted Sens.io, SensCrypt reduces
the upload operation of the \textit{real} Fitbit equipment (1481ms on average)
by 43\%.

\begin{table}
\centering
\begin{tabular}{l r r r}
\toprule
\textbf{Solutions} & \textbf{T} & \textbf{W} & \textbf{Comm}\\
\midrule
SensCrypt & 502.13 & 190.4 & 153\\
FitCrypt (Fitbit) & 904.56 & 177.36 & 162\\
FitCrypt (Garmin) & 9366 & 322 & 1686\\
\bottomrule
\end{tabular}
\caption{Upload: comparison of tracker, webserver and
communication delays (shown in ms) of SensCrypt and FitCrypt. FitCrypt (RSA or
ECC) is shown both for the Fitbit (96KB) and Garmin (1MB) memory size. The
delay of SensCrypt is independent of $mem$ size, and significantly shorter.
}
\label{table:upload}
\end{table}


\subsection{Battery Impact}

To evaluate the impact of SensCrypt on the battery lifetime, we powered the
Sens.io device using a 9V alkaline battery~\cite{duracell}. In a first
experiment, we evaluated the ability of SensCrypt to mitigate the effects of
the battery drain attack.  For this, we used the Bluetooth enabled Sens.io
device to establish a connection with an Android app running on a Nexus 4 base.
We investigated and compared two scenarios. In the first scenario, the
Bluetooth enabled Sens.io runs the Fitbit protocol to process and respond to
requests received every 15s.  In the second scenario, the Sens.io device runs
SensCrypt to process the same requests. Each scenario is performed using a
fresh 9V battery.

When running Fitbit, the Sens.io device runs out of battery after 484 minutes.
When running SensCrypt, the Sens.io device lasts for a total of 821 minutes.
Thus, SensCrypt extends the battery lifetime of Sens.io under the battery
depletion attack by 69\%.

\begin{figure}
\begin{center}
\includegraphics[width=2.9in]{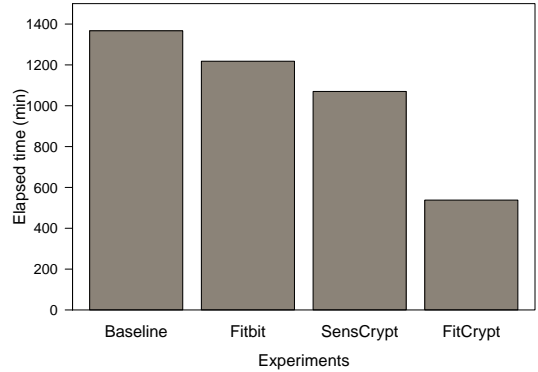}
\caption{Battery lifetime for 9V cell powered Sens.io device in four
scenarios: Baseline, Fitbit, SensCrypt and FitCrypt-RSA-256. The last three
scenarios record sensor data every 2s. The Baseline scenario measures the
battery lifetime of Arduino device with no functionality. SensCrypt reduces
13\% of the battery lifetime over Fitbit's operation. Even a vulnerable
FitCrypt-RSA-256 reduces the battery lifetime to half of SensCrypt.}
\label{fig:battery}
\end{center}
\end{figure}

In a second experiment, we compared the impact of the periodic SensCrypt,
FitCrypt-RSA-256 and Fitbit sensor data recording operations on the Sens.io
battery lifetime. In the experiment, we considered a 2s interval between
consecutive sensor recording operations. We have tested several RSA key sizes
(2048 to 256 bit long). An (insecure) RSA key size of 256 bits was the largest
value that did not hang on an Arduino board after only a few encryptions.  We
have also run a baseline experiment, measuring the battery lifetime of an
Arduino board that is not recording any sensor data.

Figure~\ref{fig:battery} shows our results. In the Baseline scenario, the
battery lasted 56 hrs and 23 mins.  When running Fitbit's sensor data record
operation, the battery lasted 50 hrs and 18 mins. When running SensCrypt's
$RecordData$ operation, the battery lasted 43 hrs and 38 mins. Thus, Fitbit's
sensor recording operation shortens the battery by 10\% over the baseline.
SensCrypt's $RecordData$ reduces the battery lifetime by 13\% of the Fitbit
battery lifetime. Finally, when running FitCrypt-RSA-256, the battery lasted
only 22 hrs and 10 mins. Even with a vulnerable key size, FitCrypt reduces the
battery lifetime by 49\% of the SensCrypt lifetime. This confirms the
unsuitability of public key cryptosystems to secure resource constrained
fitness trackers.

\section{Related Work}
\label{sec:related}

In the context of implantable medical devices (IMDs) Halperin et
al.~\cite{HHBRCDMFKM08} introduce novel software radio attacks and propose zero
power notification, authentication and key exchange solutions. Rasmussen et
al.~\cite{RCHBC09} propose proximity based access control solutions for IMDs.
The different mission of fitness trackers creates different design constraints.
First, unlike IMD security, where the focus is on authentication and key
exchange, SensCrypt's focus is on the secure storage and communication of
tracker data. This is further emphasized by our need to also consider attackers
that can perform Capture and JTAG-R attacks, for both trackers and bases
(readers in the IMD context).  While such attacks may not be possible for IMDs,
and IMD readers may be expensive enough to afford tamper proof memory, these
assumptions do not hold for most existing fitness centric social sensor network
solutions. Furthermore, while additional user interaction may be naturally
accepted for IMDs, fitness security solutions should minimize or even eliminate
user involvement.

Tsubouchi et al.~\cite{TKS13} have shown that Fitbit data can be used to infer
surprising information, in the form of working relations between tracker
carrying co-workers.  This information could be used to surreptitiously learn
the organizational profile of a company. This work assumes access to the
fitness data of other users, a task that part of our paper undertakes.

Naveed et al.~\cite{NZDWG14} introduced an ``external device mis-bonding
attack'' for Bluetooth enabled Android health/medical devices, then collected
sensitive user data from and fed arbitrary information into the user's account.
They developed Dabinder, an OS level defense that generates and enforces secure
bonding policies between a device and its official app. Our work differs in the
types and implementation of attacks, and in the solution placement: SensCrypt
is implemented at the tracker and webserver, whereas Dabinder is focused on the
base.

Lim et al.~\cite{WBAN} analyzed the security of a remote cardiac monitoring
system where the data originating from the sensors is sent through a Body Area
Network (BAN) gateway and a wireless router to a final monitoring server.
Muraleedharan et al.~\cite{Muraleedharan} proposed DoS attacks including
Sybil~\cite{Sybil} and wormhole~\cite{Wormhole} attacks, for a health
monitoring system using wireless sensor networks. They introduced an
energy-efficient cognitive routing algorithm to address such attacks.
Our work differs through its system architecture,
communication model and tracker capabilities.

Barnickel et al.~\cite{HealthNet} targeted security and privacy issues for
HealthNet, a health monitoring and recording system. They proposed a security
and privacy aware architecture, relying on data avoidance, data minimization,
decentralized storage, and the use of cryptography.
Marti et al.~\cite{Ramon}
described the requirements and implementation of the security mechanisms for
MobiHealth, a wireless mobile health care system. MobiHealth relies on
Bluetooth and ZigBee link layer security for communication to the sensors
and uses HTTPS mutual authentication and encryption for connections to the
backend.

\section{Conclusions}
\label{sec:conclusions}

We identified and exploited vulnerabilities in the design of Fitbit and Garmin,
to launch inspection and injection attacks. We presented SensCrypt, a secure
and efficient solution for storing and communicating tracker sensor data.
SensCrypt imposes minimal computation and communication overhead on
trackers, and is resilient even to attackers able to probe the memory of
captured trackers.

\section{Acknowledgments}

\noindent
This research was supported in part by DoD grant W911NF-13-1-0142 and NSF grant
EAGER-1450619.

\begin{IEEEbiography}
[{\includegraphics[width=1.07in,height=1in,clip,keepaspectratio]{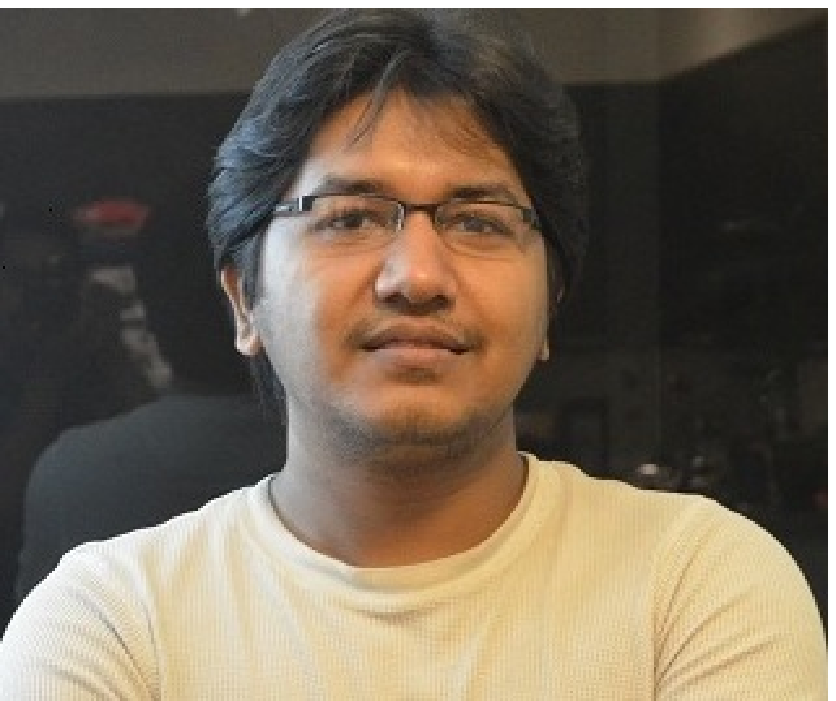}}]
{Mahmudur Rahman} is a Ph.D candidate in the School of Computing and
Information Sciences at Florida International University. He received his
Bachelor's degree in C.S.E from BUET, Bangladesh. His research interests
include security and privacy with applications in online and geosocial
networks, wireless networks, distributed computing systems and mobile
applications.
\end{IEEEbiography}

\begin{IEEEbiography}
[{\includegraphics[width=1.07in,height=1in,clip,keepaspectratio]{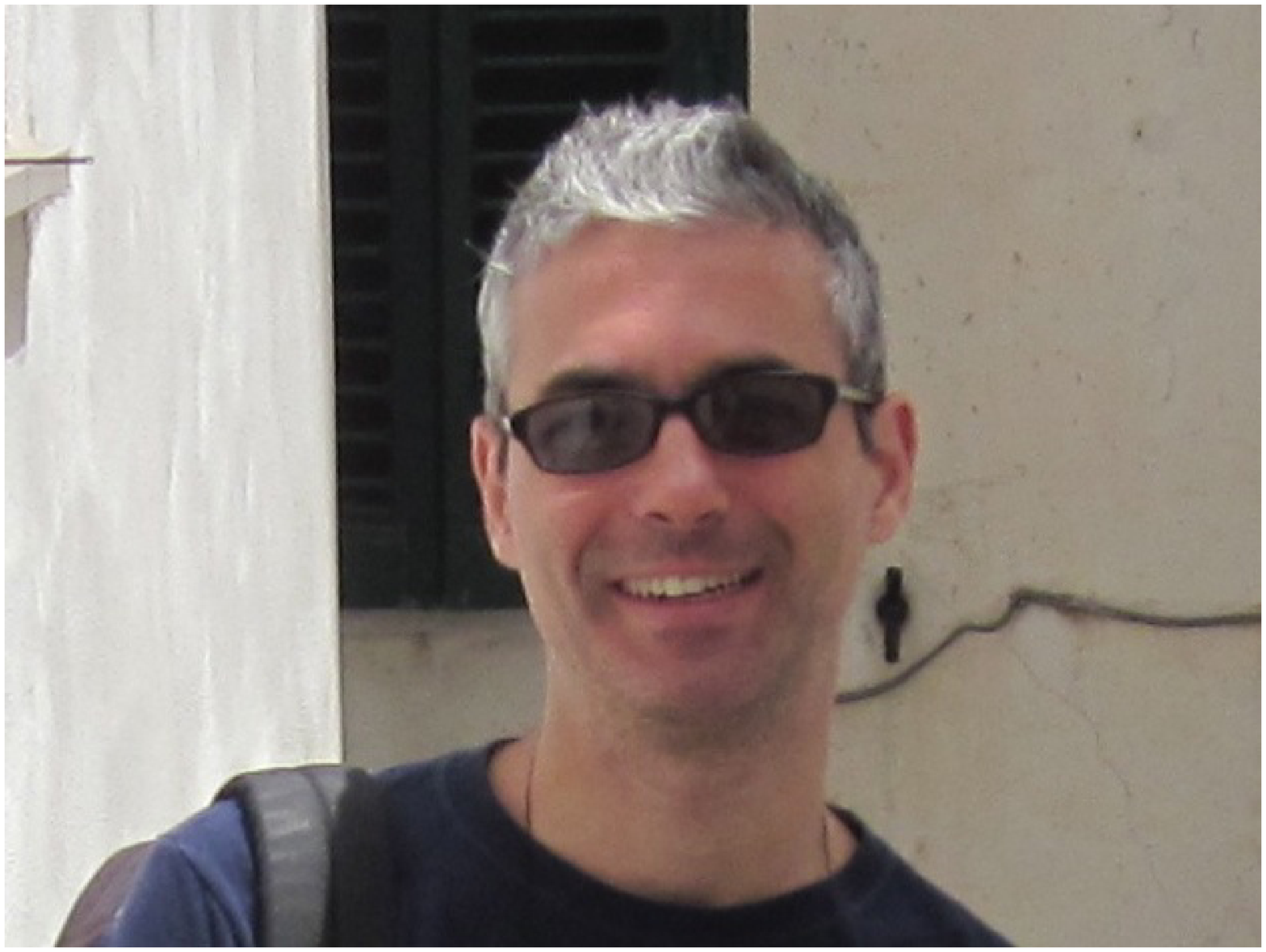}}]
{Bogdan Carbunar} is an assistant professor in SCIS at FIU. Previously, he held
various researcher positions within the Applied Research Center at Motorola.
His research interests include distributed systems, security and applied
cryptography.  He holds a Ph.D. in Computer Science from Purdue University.
\end{IEEEbiography}

\begin{IEEEbiography}
[{\includegraphics[width=1.07in,height=1in,clip,keepaspectratio]{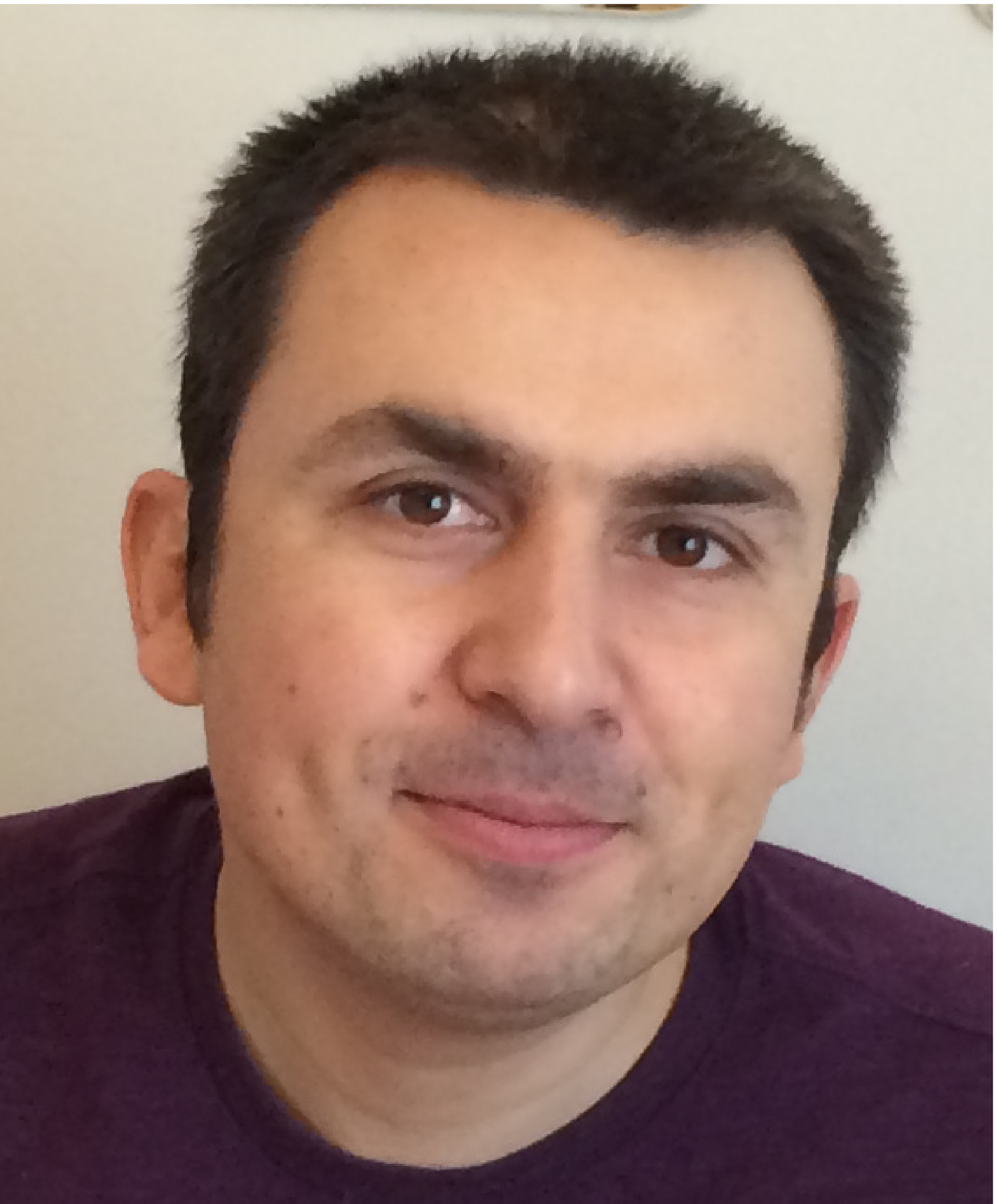}}]
{Umut Topkara} is a research engineer with JW Player. Previously, he held research
and engineering positions at the IBM T.J. Watson Research Lab and Google. His 
research interests include mobile collaboration, security, and machine learning. 
He holds a Ph.D. in Computer Science from Purdue University.
\end{IEEEbiography}


\begin{thebibliography}{10}

\bibitem{HolterM}
{Holter Monitor}.
\newblock \url{https://en.wikipedia.org/wiki/Holter_monitor}.

\bibitem{Nike+}
{Nike+}.
\newblock \url{http://nikeplus.nike.com/plus/}.

\bibitem{Fitbit}
Fitbit.
\newblock \url{http://fitbit.com/}.

\bibitem{Forerunner}
{Garmin Forerunner}.
\newblock \url{http://sites.garmin.com/forerunner610/}.

\bibitem{Jawbone}
{Jawbone UP24}.
\newblock \url{https://jawbone.com/up}.

\bibitem{BodyMedia}
{Body Media}.
\newblock \url{http://www.bodymedia.com/}.

\bibitem{BMData}
{Jawbone takes a big bite out of health tech: acquires BodyMedia, launches Up
  app platform}.
\newblock
  http://venturebeat.com/2013/04/30/jawbone-takes-a-big-bite-out-of-health-tec%
h-acquires-bodymedia-launches-up-app-platform.

\bibitem{PleaseRobMe}
{Please Rob Me}.
\newblock \url{http://www.http://pleaserobme.com/}.

\bibitem{TKS13}
Kota Tsubouchi, Ryoma Kawajiri, and Masamichi Shimosaka.
\newblock Working-relationship detection from fitbit sensor data.
\newblock In {\em Proceedings of the 2013 ACM conference on Pervasive and
  ubiquitous computing adjunct publication}, UbiComp '13 Adjunct, pages
  115--118, 2013.

\bibitem{HHBRCDMFKM08}
D.~Halperin, T.~Heydt-Benjamin, B.~Ransford, S.~Clark, B.~Defend, W.~Morgan,
  K.~Fu, T.~Kohno, and W.~Maisel.
\newblock Pacemakers and implantable cardiac defibrillators: Software radio
  attacks and zero-power defenses.
\newblock In {\em Proceedings of IEEE Symposium on Security and Privacy}, pages
  129--142, 2008.

\bibitem{Insulin}
Chunxiao Li, A.~Raghunathan, and N.K. Jha.
\newblock Hijacking an insulin pump: Security attacks and defenses for a
  diabetes therapy system.
\newblock In {\em Proceedings of the IEEE International Conference on e-Health
  Networking Applications and Services (Healthcom)}, 2011.

\bibitem{RCHBC09}
K.~B. Rasmussen, C.~Castelluccia, T.~S. Heydt-Benjamin, and S.~Capkun.
\newblock Proximity-based access control for implantable medical devices.
\newblock In {\em ACM Conference on Computer and Communications Security},
  2009.

\bibitem{NZDWG14}
Muhammad Naveed, Xiaoyong Zhou, Soteris Demetriou, XiaoFeng Wang, and Carl~A
  Gunter.
\newblock Inside job: Understanding and mitigating the threat of external
  device mis-bonding on android.
\newblock In {\em Proceedings of ISOC Network and Distributed Computing
  Security (NDSS)}, 2014.

\bibitem{MotoActv}
{Mototola MotoActv}.
\newblock
  \url{http://www.motorola.com/us/MOTOACTV-16GB-Golf-Edition/121481.html}.

\bibitem{Basis}
{Basis B1}.
\newblock \url{http://www.mybasis.com/}.

\bibitem{Nest}
{Nest Thermostat}.
\newblock \url{https://nest.com/thermostat/life-with-nest-thermostat/}.

\bibitem{WeMo}
{WeMo Switch}.
\newblock \url{http://www.belkin.com/us/p/P-F7C027/}.

\bibitem{Mother}
{Sense: The meaning of life}.
\newblock \url{https://sen.se/store/mother/}.

\bibitem{Nike}
{Nike+ Fuelband SE}.
\newblock \url{https://secure-nikeplus.nike.com/plus/}.

\bibitem{BMR}
A.~J. Hulbert and P.~L. Else.
\newblock {Basal Metabolic Rate: History, Composition, Regulation, and
  Usefulness}.
\newblock {\em Physiological and Biochemical Zoology}, 77(6):869--876, 2004.

\bibitem{Firstbeat}
Vo2 estimation method based on heart rate measurement.
\newblock Technical report, Firstbeat Technologies Ltd, 2005.

\bibitem{B06}
Ing Breeuwsma.
\newblock {Forensic imaging of embedded systems using JTAG (boundary-scan)}.
\newblock {\em Digital Investigation}, 3, 2006.

\bibitem{Reverse.engineer}
{17 U.S. Code § 1201 - Circumvention of copyright protection systems}.
\newblock \url{https://www.law.cornell.edu/uscode/text/17/1201}.

\bibitem{libfitbit}
Libfitbit: Library for accessing and transfering data from the fitbit health
  device.
\newblock \url{https://github.com/qdot/libfitbit}.

\bibitem{ANT-FS}
{ANT-FS and FIT}.
\newblock \url{http://www.thisisant.com/developer/ant/ant-fs-and-fit}.

\bibitem{FitbitSpecs}
{Fitbit Specs}.
\newblock \url{http://www.fitbit.com/one/specs}, Last retrieved on October 1st,
  2013.

\bibitem{HMAC}
Mihir Bellare, Ran Canetti, and Hugo Krawczyk.
\newblock Keying hash functions for message authentication.
\newblock In {\em Proceedings of the 16th Annual International Cryptology
  Conference on Advances in Cryptology}, CRYPTO '96, pages 1--15, 1996.

\bibitem{Earndit}
{Earndit: We reward you for exercising}.
\newblock \url{http://earndit.com/}.

\bibitem{TCX}
{Training Center XML (TCX)}.
\newblock \url{http://developer.garmin.com/schemas/tcx/v2/}.

\bibitem{RMMR11}
Kiran~K. Rachuri, Cecilia Mascolo, Mirco Musolesi, and Peter~J. Rentfrow.
\newblock Sociablesense: Exploring the trade-offs of adaptive sampling and
  computation offloading for social sensing.
\newblock In {\em Proceedings of the 17th Annual International Conference on
  Mobile Computing and Networking}, MobiCom '11, pages 73--84, New York, NY,
  USA, 2011. ACM.

\bibitem{arduinoUno}
{Arduino Uno}.
\newblock \url{http://arduino.cc/en/Main/arduinoBoardUno}.

\bibitem{ArduinoGuide}
{Arduino Guide}.
\newblock \url{http://arduino.cc/en/Guide/Introduction}.

\bibitem{Bluetooth}
Bluetooth SIG.
\newblock Specification of the bluetooth system, 2001.

\bibitem{GPWES04}
Nils Gura, Arun Patel, Arvinderpal Wander, Hans Eberle, and Sheueling~Chang
  Shantz.
\newblock {Comparing Elliptic Curve Cryptography and RSA on 8-bit CPUs}.
\newblock In {\em Proceedings of Cryptographic Hardware and Embedded Systems
  (CHES)}, pages 119--132, 2004.

\bibitem{duracell}
{Duracell Product Data Sheets}.
\newblock
  \url{ww2.duracell.com/media/en-US/pdf/gtcl/Product_Data_Sheet/NA_DATASHEETS/%
PC1604_US_PC.pdf}.

\bibitem{WBAN}
S.~Lim, T.H. Oh, Y.~Choi, and T.~Lakshman.
\newblock Security issues on wireless body area network for remote healthcare
  monitoring.
\newblock In {\em Proceedings of the IEEE International Conference on Sensor
  Networks, Ubiquitous, and Trustworthy Computing (SUTC)}, pages 327--332,
  2010.

\bibitem{Muraleedharan}
Rajani Muraleedharan and Lisa~Ann Osadciw.
\newblock Secure health monitoring network against denial-of-service attacks
  using cognitive intelligence.
\newblock In {\em Proceedings of the Communication Networks and Services
  Research Conference}, pages 165--170, 2008.

\bibitem{Sybil}
J.~Newsome, E.~Shi, D.~Song, and A.Perrig.
\newblock The sybil attack in sensor networks: Analysis and defenses.
\newblock In {\em Third International Symposium on Information Processing in
  Sensor Networks(IPSN)}, 2004.

\bibitem{Wormhole}
C.~Karlof and D.Wagner, editors.
\newblock {\em Secure Routing in Sensor Networks: Attacks and Countermeasures},
  2003.

\bibitem{HealthNet}
Johannes Barnickel, Hakan Karahan, and Ulrike Meyer.
\newblock Security and privacy for mobile electronic health monitoring and
  recording systems.
\newblock In {\em Proceedings of the IEEE International Symposium on A World of
  Wireless, Mobile and Multimedia Networks (WoWMoM)}, pages 1--6, 2010.

\bibitem{Ramon}
Ramon Marti and Jaime Delgado.
\newblock Security in a wireless mobile health care system, 2007.

\end{thebibliography}
\end{document}